\documentclass[prd,amsmath,preprintnumbers,floatfix,twocolumn,letterpaper,showpacs]{revtex4}

\usepackage{graphicx}
\usepackage{slashed}

\newcommand{\nc}{\newcommand}

\nc{\eref}[1]{~(\ref{#1})}
\nc{\etal}{\emph{~et al.}}
\nc{\ecite}[1]{~\cite{#1}}
\nc{\efig}[1]{~Fig.~\ref{#1}}
\nc{\etable}[1]{~Table~\ref{#1}}

\begin{document}

\preprint{}

\title{Quark Propagation in the Instantons of Lattice QCD}

\author{Daniel Trewartha}
\email{daniel.trewartha@adelaide.edu.au}
\author{Waseem Kamleh}
\author{Derek Leinweber}
\author{Dale S. Roberts}
\address{Centre for the Subatomic Structure of Matter, School of Chemistry and Physics, University of Adelaide, SA, 5005, Australia}



\begin{abstract}
We quantitatively examine the extent to which instanton degress of freedom, contained within standard Monte-carlo generated gauge-field configurations, can maintain the characteristic features of the mass and renormalisation functions of the non-perturbative quark propagator. We use over-improved stout-link smearing to isolate instanton effects on the lattice. Using a variety of measures, we illustrate how gauge fields consisting almost solely of instanton-like objects are produced after only 50 sweeps of smearing. We find a full vacuum, with a packing fraction more than three times larger than phenomenological models predict. We calculate the overlap quark propagator on these smeared configurations, and find that even at high levels of smearing the majority of the characteristic features of the propagator are reproduced. We thus conclude that instantons contained within standard Monte-carlo generated gauge-field configurations are the degrees of freedom responsible for the dynamical generation of mass observed in lattice QCD.
\end{abstract}

\maketitle

\section{Introduction}
Instantons are believed to be an essential component of the long-distance physics of the QCD vacuum, and the lattice provides a unique opportunity to gain insight into their role. In this study, we will for the first time quantitatively examine the extent to which instanton degrees of freedom, contained within standard Monte-Carlo generated gauge-field configurations, can maintain the characteristic features of the mass and renormalisation functions of the non-perturbative quark propagator.\par 
In order to isolate the effects of instanton degrees of freedom a UV filter is required to remove topologically non-trivial fluctuations. A variety of filters have been used, including cooling\cite{Berg:1981nw,Teper:1985rb,Ilgenfritz:1985dz}, APE smearing\cite{Albanese:1987ds}, HYP smearing \cite{Hasenfratz:2001hp} and stout link smearing \cite{Morningstar:2003gk}, among others. These algorithms can suffer from destruction of  the instanton content of the vacuum, and so in this work we use over-improved stout-link smearing \cite{Garcia Perez:1993ki,Moran:2008ra}, a form of smearing tuned to preserve instantons. Section \ref{sec:smearing} briefly describes these smearing methods.\par 
We then seek to quantify the effects of smearing on the lattice gauge fields in section \ref{sec:smearingeffect}. We produce configurations dominated by instanton-like objects, and compare to the phenomenological instanton liquid model \cite{Shuryak:1981ff,Shuryak:1982dp,Shuryak:1982hk}, which models the vacuum as composed of a constant number of instantons and anti-instantons of constant radius. \par 
We will then briefly introduce the Fat Link Irrelevant Clover(FLIC) overlap action \cite{Kamleh:2001ff} in section \ref{sec:overlap}, an improved fermion action with a lattice deformed version of chiral symmetry, which removes the problem of additive mass renormalisation of the quark propagator. Results will be compared for smeared and unsmeared configurations in section \ref{sec:results}, and conclusions summarised in section \ref{sec:conclusion}.
\section{Over-Improved Stout-Link Smearing}
\label{sec:smearing}
It has long been known \cite{Teper:1985rb,Ilgenfritz:1985dz,Kovacs:1999iq} that smearing the lattice reveals objects which approximate classical instantons. However, it has been known for nearly as long that excessive smearing can destroy the same instanton-like objects or distort their structure.
\par 
We can understand this behaviour by explicitly considering the single instanton solution, \cite{Belavin:1975fg,Kusterer:2001vk,'tHooft:1976up,'tHooft:1976fv}
\begin{eqnarray}
\label{instsol}
A_{\mu}(x,x_{0}) &=& \frac{i}{g}\frac{(x-x_{0})^{2}}{(x-x_{0})^{2}+\rho^{2}}(\delta_{\mu}S)S^{-1}, \nonumber \\
S &=& \frac{x_{4} \pm i\vec{x} . \vec{\sigma}}{\sqrt{x^{2}}}\mathrm{,}
\end{eqnarray}
where $\rho$ is the instanton radius, $\sigma$ the Pauli matrices and $x_{0}$ the center of the instanton.
The Wilson gauge action can be expanded in powers of $a$ as\cite{Garcia Perez:1993ki}
\begin{eqnarray}
\label{wilsonactexp}
S_{W} &=& \sum_{x,\mu ,\nu}\mathrm{Tr}[-\frac{a^{4}}{2}F_{\mu \nu}^{2} + \frac{a^{6}}{24} (({\bf D}_{\mu}F_{\mu \nu}(x))^{2} + ({\bf D}_{\nu}F_{\mu \nu}(x))^{2}) \nonumber \\
&\quad& - \frac{a^{8}}{24}\lbrace F_{\mu \nu}^{4}(x) + \frac{1}{30}(({\bf D}_{\mu}^{2}F_{\mu \nu}(x))^{2} + ({\bf D}^{2}_{\nu}F_{\mu \nu}(x))^{2})\nonumber \\
&\quad&  + \frac{1}{3}{\bf D}_{\mu}^{2}F_{\mu \nu}(x){\bf D}_{\nu}^{2}F_{\mu \nu}(x) - \frac{1}{4}({\bf D}_{\mu}{\bf D}_{\nu}F_{\mu \nu}(x))^{2}\rbrace]\mathrm{,} \nonumber \\
\quad 
\end{eqnarray}
where ${\bf D}_{\mu} f = [D_{\mu},f]$ for arbitrary $f$.\par 
Inserting the instanton solution of Eq.~(\ref{instsol}) into Eq.~(\ref{wilsonactexp}), we acquire
\begin{equation}
S^{inst}_{w}=\frac{8 \pi^{2}}{g^{2}}[1 - \frac{1}{5}(\frac{a}{\rho})^{2} - \frac{1}{70}(\frac{a}{\rho})^{4} + \mathcal{O}(a^{6})].
\end{equation}
The source of the problem is clear; although at first order this is equal to the continuum instanton action, the leading order error term is strictly negative and inversely proportional to $\rho$. The use of a plaquette-based smearing algorithm will reduce the size of instantons in decreasing the action and ultimately enable lattice artifacts to spoil and remove the instanton. This situation is not ameliorated by using improved actions such as the Symanzik action \cite{Garcia Perez:1993ki}\par 
This issue can be mitigated using over-improved stout-link smearing, a form of smearing designed to preserve instantons. This was first implemented in Ref.~\cite{Garcia Perez:1993ki}, although here we follow the work Ref.~\cite{Moran:2008ra}, which uses a slightly modified combination of links. One introduces a new parameter, $\epsilon$, defining
\begin{equation}
S(\epsilon) = \frac{2}{g^{2}} \sum_{x} \sum_{\mu > \nu}[\frac{5-2\epsilon}{3}(1-P_{\mu\nu}) - \frac{1-\epsilon}{12}(1-R_{\mu\nu})],
\end{equation}
with $P_{\mu\nu}$ and $R_{\mu\nu}$ the $1 \times 1$ and $1 \times 2$ plaquettes. We note that a value of $\epsilon = 1$ gives the Wilson plaquette action and $\epsilon = 0$ gives the Symanzik improved action \cite{Symanzik:1983dc}. Again substituting the instanton solution into this, one acquires \cite{Moran:2008ra}
\begin{equation}
\label{eq:SLsmac}
S^{inst}(\epsilon) = \frac{8 \pi^{2}}{g^{2}}[1 - \frac{\epsilon}{5}(\frac{a}{\rho})^{2} + \frac{14 \epsilon - 17}{210}(\frac{a}{\rho})^{4}].
\end{equation}
A negative value of $\epsilon$ will preserve instanton-like objects by making the first order error term positive. This has however, simultaneously introduced a new problem; we have removed the possibility of shrinking objects and replaced it with that of enlarging, and so smearing can distort the topological structure of the lattice if used excessively with large negative values of $\epsilon$. Following Ref.~\cite{Moran:2008ra}, we adopt the small value $\epsilon=-0.25$.
\par
\begin{figure}[h!!]
\includegraphics[trim=3cm 23cm 10cm 3cm, clip=true,width=\hsize]{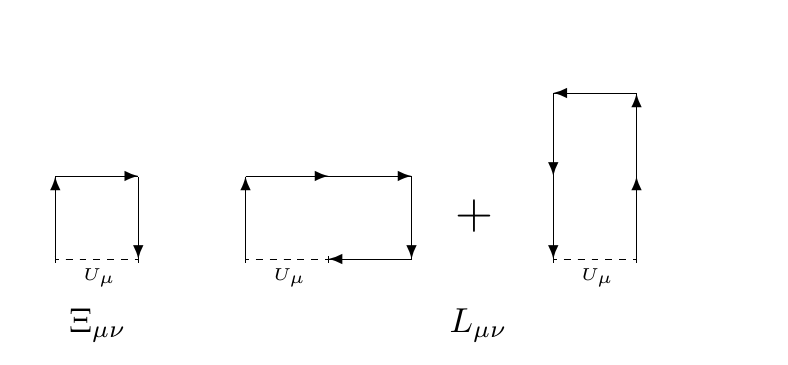}
\caption{Components of $\Xi_{\mu\nu}$ and $L_{\mu\nu}$}
\label{pic:stapledef}
\end{figure}

 
Explicitly, a sweep of over-improved stout-link smearing is implemented by replacing all links on the lattice with a smeared link, defined by
\begin{equation}
U'_{\mu}(x)=\exp(iQ_{\mu}(x))U_{\mu}(x),
\end{equation}
where
\begin{equation}
Q_{\mu}(x)=\frac{i}{2}(\Omega_{\mu}^{\dagger}(x) - \Omega_{\mu}(x)) - \frac{i}{2N}\mathrm{Tr}(\Omega_{\mu}^{\dagger}(x) - \Omega_{\mu}(x)),
\end{equation}
with
\begin{equation}
\Omega_{\mu}(x)=C_{\mu}(x)U_{\mu}^{\dagger}(x).
\end{equation}
Then we define
\begin{eqnarray}
C_{\mu}(x)=\rho\sum_{\nu}\left[\frac{5-2\epsilon}{3}(\Xi_{\mu\nu})(x) - \frac{1-\epsilon}{12}(L_{\mu\nu}(x))\right],
\end{eqnarray}
where $\rho$ is a parameter controlling the level of smearing and $\Xi_{\mu\nu}$ are 'staples'; the 3 links in the $\mu \nu$ plane forming the $P_{\mu\nu}$ plaquette with $U_{\mu}$ removed, and $L_{\mu\nu}$ are analogously defined for $2 \times 1$ rectangles, illustrated in Fig.~\ref{pic:stapledef}.
\par
Following the work in Ref.~(\cite{Moran:2008ra}), we choose $\rho = 0.06$ and $\epsilon=-0.25$, as these provide values of $S(\epsilon)/S_{continuum}$ close to 1, and preserve instanton-like objects on the lattice with size above the dislocation threshold of $1.97 \, a$. We note however, that small objects can still be destroyed by the smearing process. Pair annihilation can also remove them from the lattice.
\section{Effects of Smearing}
\label{sec:smearingeffect}
We wish to find a smearing level such that the configurations are dominated by topological objects as similar to continuum instantons as possible. At the same time, one needs to be wary of distorting their topological structure by enlarging or pair annihilating these objects. We will quantify the effects of smearing in order to choose an optimal balance between these two effects.\par 
The work in Ref.~\cite{Moran:2008qd} suggests that after just 20 sweeps of over-improved stout link smearing, topological objects found on the lattice closely approximate instantons. Here we adopt a similar approach to that taken in Ref.~\cite{Moran:2008qd}, searching the lattice for sites which are local maxima of the action in their surrounding hypercube \cite{Kusterer:2001vk}. These are then taken as the approximate centre of an (anti-)instanton, around which we fit the classical instanton action density,
\begin{equation}
\label{instactdens}
S_{0}(x) = \xi \frac{6}{\pi^{2}}\frac{\rho^{4}}{((x-x_{0})^{2}+\rho^{2})^{4}},
\end{equation}
where $\xi$, $\rho$ and $x_{0}$ are fit parameters, noting that $x_{0}$ is not restricted to lattice sites. The parameter $\xi$ is introduced as lattice topological objects are expected to have a higher action than classical instantons. We wish to determine the fit by using the shape of the action density around a local maximum, rather than the height. We can then compare data obtained for the radii, $\rho$, of instanton candidates from this to the relationship between the radius and topological charge at the centre of an (anti-)instanton,
\begin{equation}
\label{topcharge}
q(x_{0})=Q\frac{6}{\pi^{2} \rho^{4}},
\end{equation}
where $Q=\mp 1$ for an (anti-)instanton. $q(x_{0})$ at the fitted values of $x_{0}$ are found using linear interpolation from neighbouring hypercubes to find an extremum inside the hypercube containing $x_{0}$.
This data will provide the basis for our investigation of the effects of smearing.\par 
\begin{figure*}
\includegraphics[trim=1cm 2cm 2cm 1.5cm, clip=true,width=0.3\hsize,angle=90]{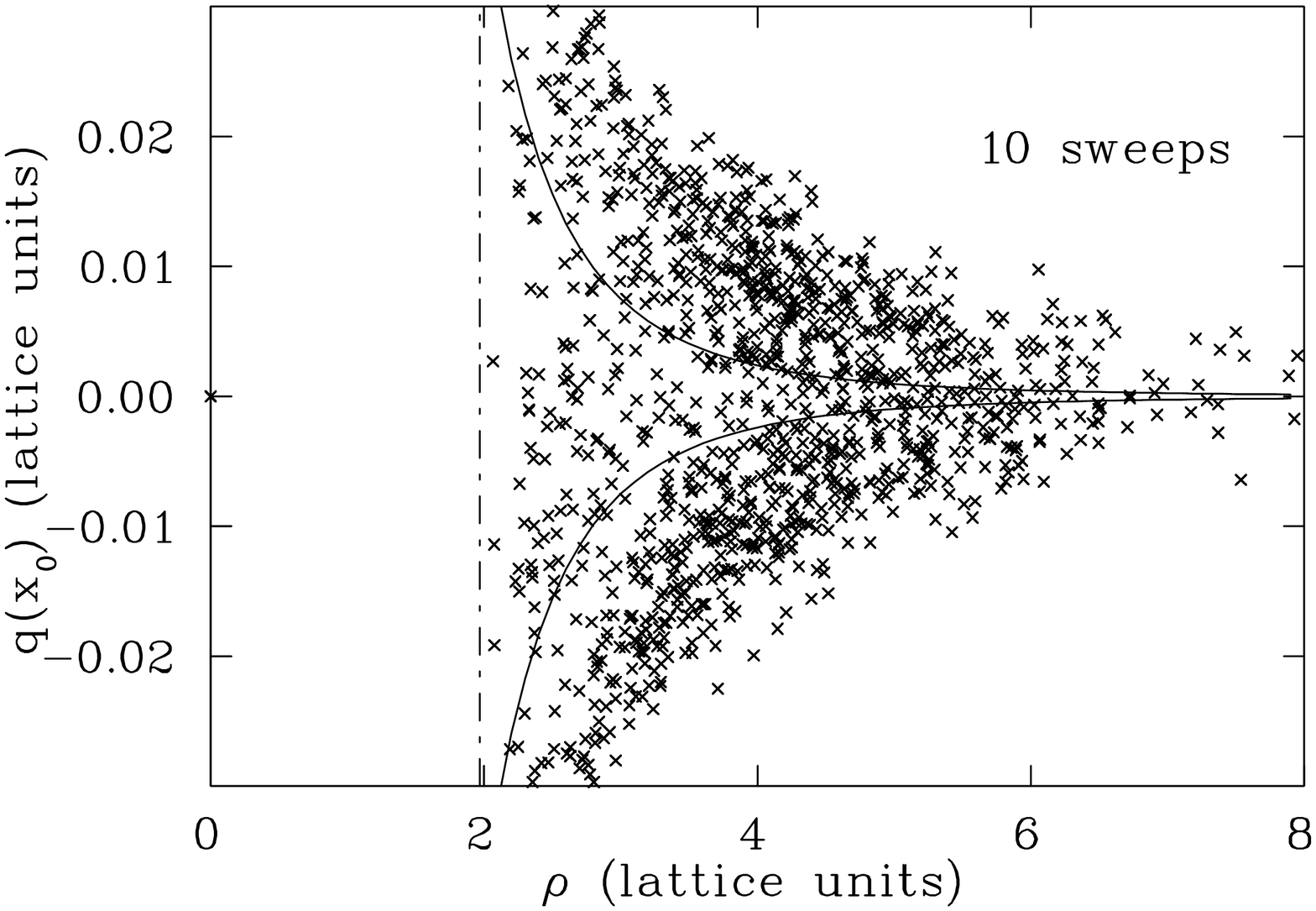}
\includegraphics[trim=1cm 2cm 2cm 1.5cm, clip=true,width=0.3\hsize,angle=90]{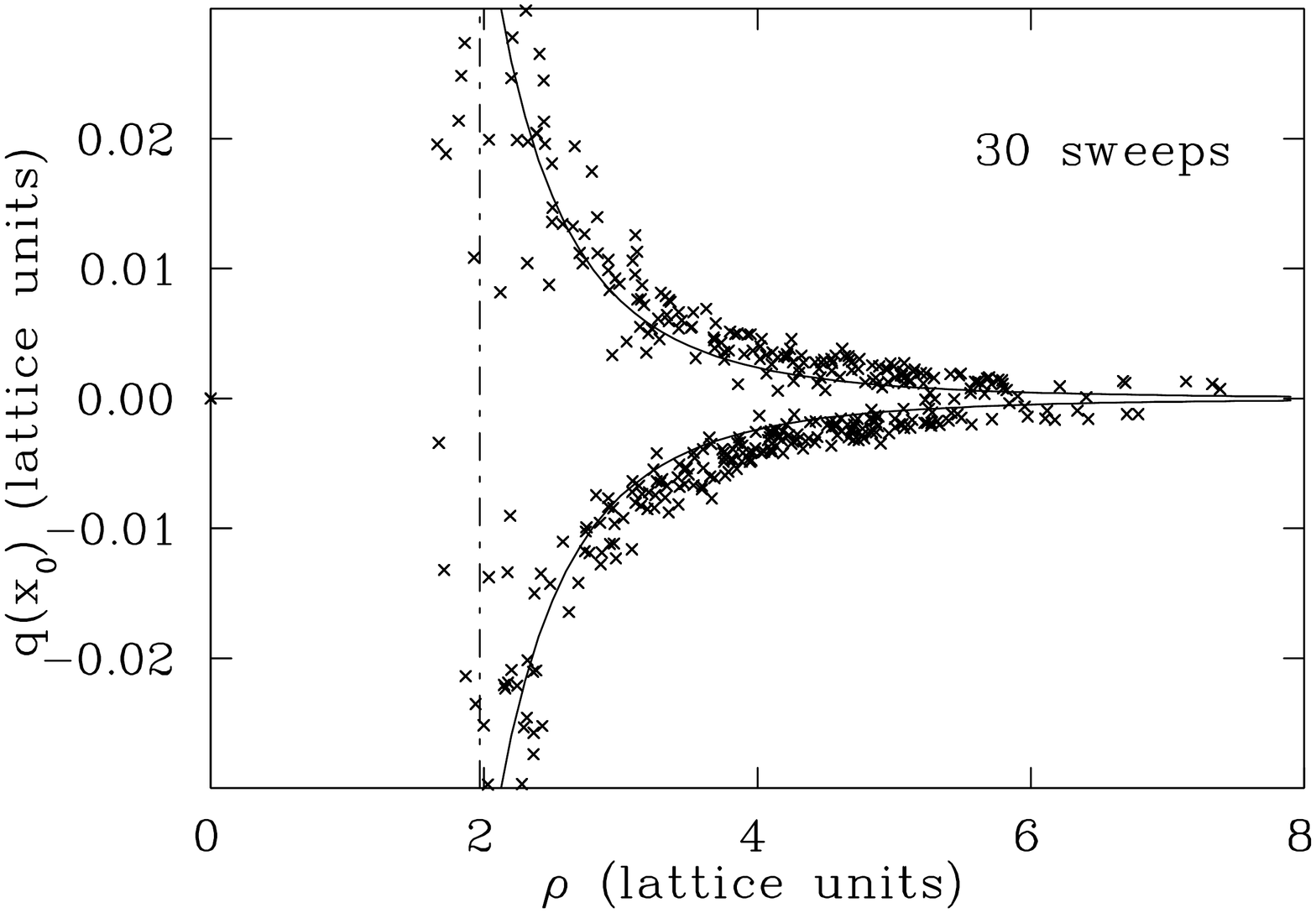}
\includegraphics[trim=1cm 2cm 2cm 1.5cm, clip=true,width=0.3\hsize,angle=90]{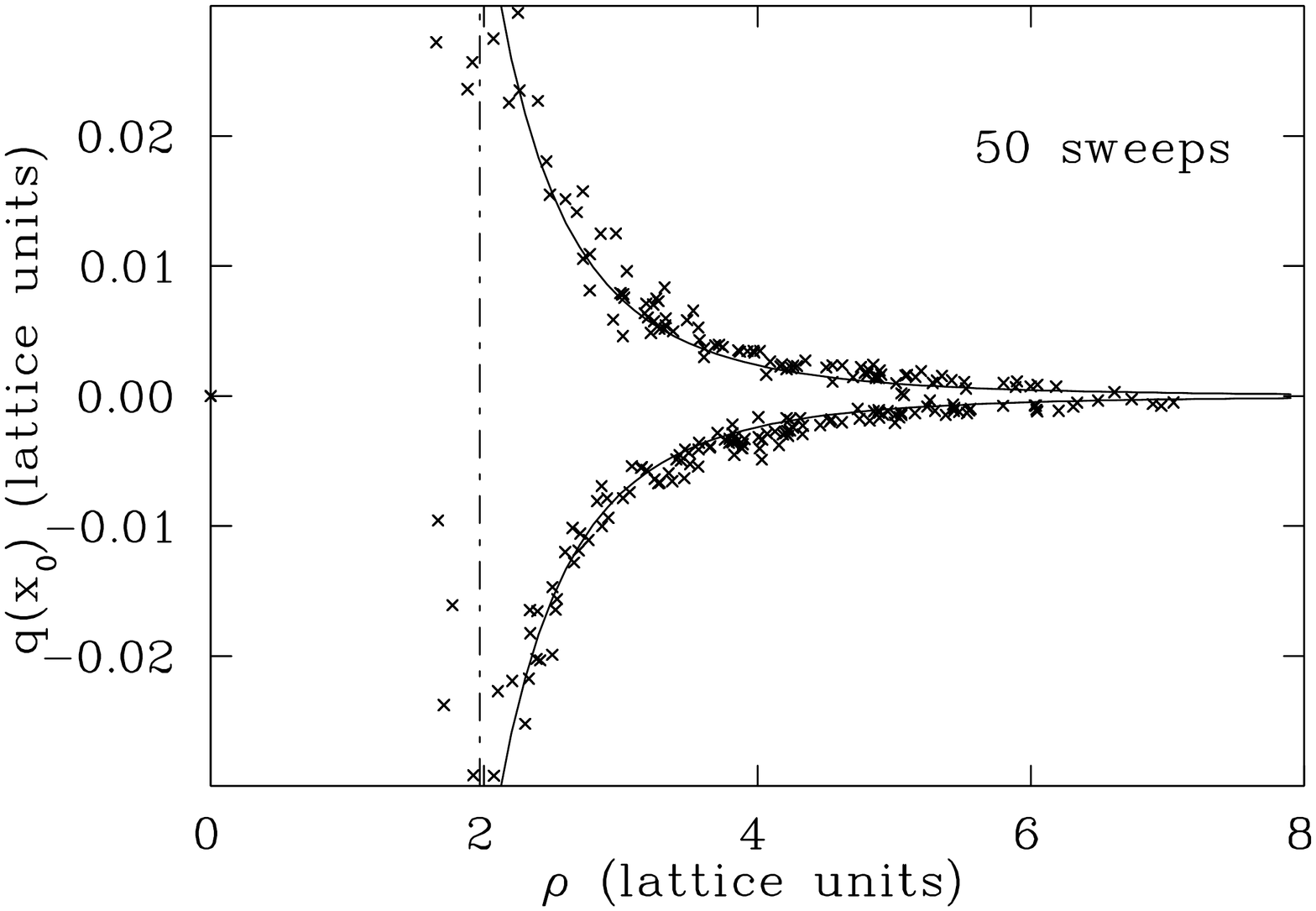}
\includegraphics[trim=1cm 2cm 2cm 1.5cm, clip=true,width=0.3\hsize,angle=90]{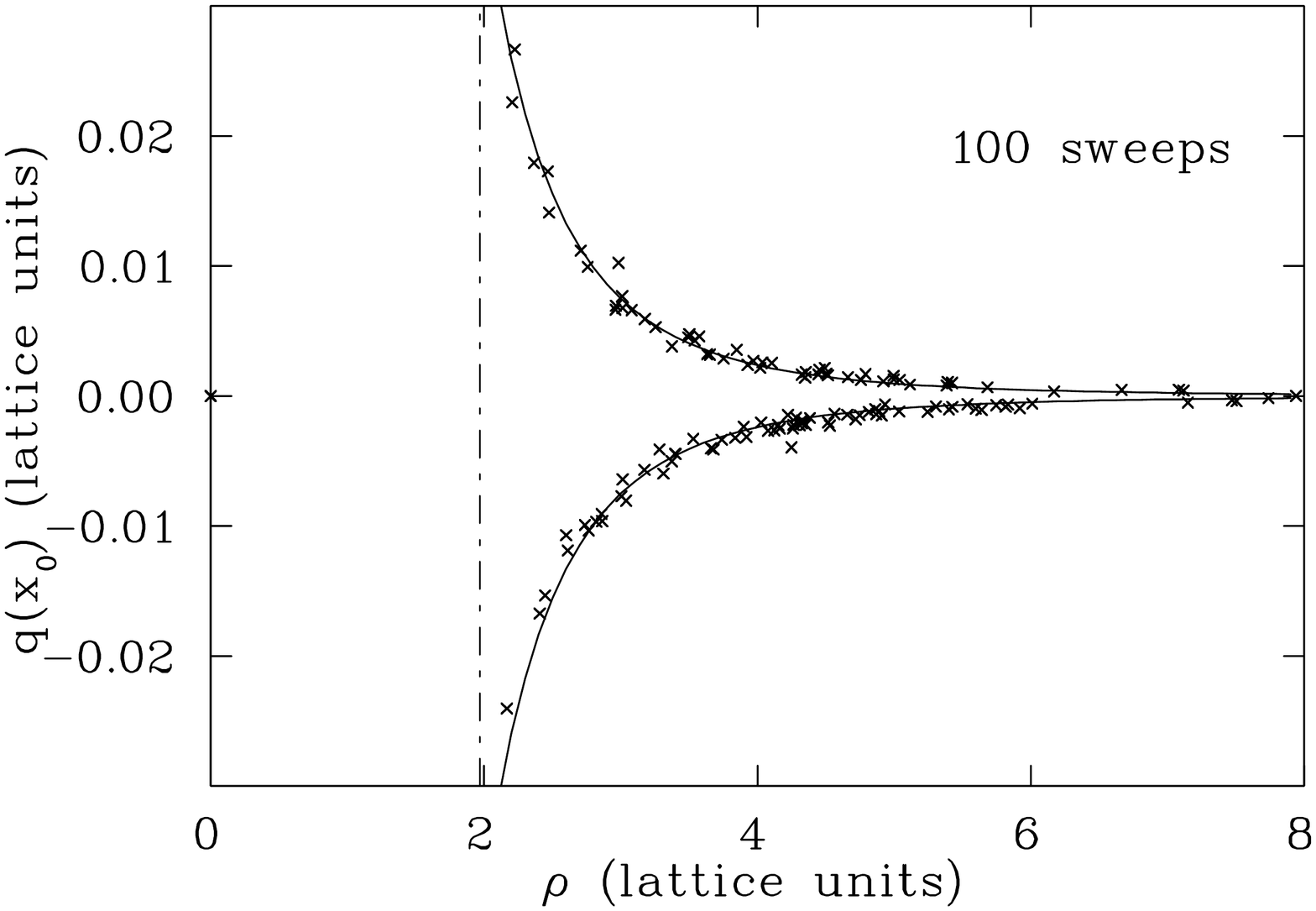}
\caption{The values of the instanton radius, $\rho$, found by fitting lattice maxima of the action to the classical instanton action density of Eq.~(\ref{instactdens}) are plotted as crosses, against the topological charge at the centre, $q(x_{0})$, on configurations with 10, 30, 50 and 100 sweeps of smearing. Results are compared to the theoretical relationship between the instanton radius and topological charge at the centre, Eq.~(\ref{topcharge}) (solid lines), and the dislocation threshold,$1.97a$ (dash-dotted line).}
\label{fig:rhovsq}
\end{figure*}
Simulations are performed on $50$ dynamical FLIC  $20^{3}\times40$ configurations, with a lattice spacing of $0.126$ fm, corresponding to a spatial extent of $2.52$ fm. Periodic boundary conditions are used. Up to 300 sweeps of smearing are investigated. To calculate the action and topological charge densities used in our fits we use the $\mathcal{O}(a^{4})$ five-loop improved action and charge densities defined in Ref.~\cite{BilsonThompson:2001ca}
\par 
It should be noted that at low levels of smearing we expect to fit a large number of false positives; local maxima of the action corresponding to noise. The degree to which fitted results concur with Eq.~(\ref{topcharge}) will thus be a key first test. This correspondence is graphed for a single configuration in Fig.~\ref{fig:rhovsq} for various smearing levels. The number of instanton candidates starts out large and distributed fairly evenly around sizes of 2-8 lattice units, with little correlation to the predicted charge lines of Eq.~(\ref{topcharge}). This quickly changes as the number of smearing sweeps increases, eventually leading to a very close fit. The number of instanton candidates also drops off rapidly at first, then steadily decreases. By the 50 sweep mark we can be confident that almost all local maxima found closely approximate an instanton near the centre. Notably, instanton candidates with low radii have systematically lower topological charge at the centre than predicted by Eq.~\ref{topcharge}. This is due to their proximity to the dislocation threshold of $1.97 a$, under which objects are shrunk.
\par
\begin{figure}
\includegraphics[trim=1cm 2cm 2cm 2.5cm, clip=true,height=\hsize,angle=90]{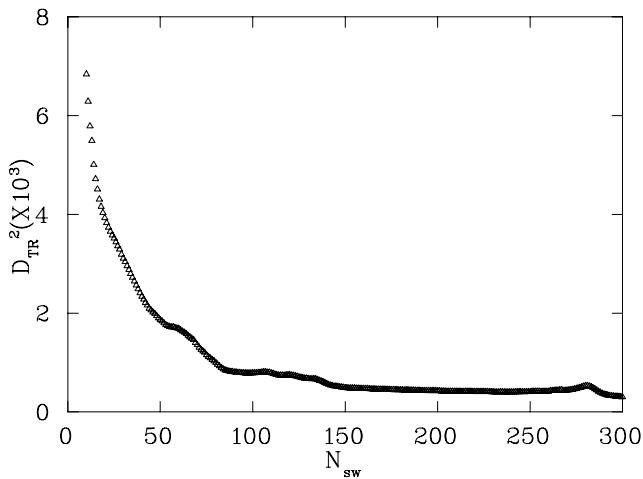}
\caption{Average squared distance from the Theoretical Relation, $D^{2}_{\textrm{TR}}$, between topological charge at the centre of an instanton candidate and its fitted radius, as a function of smearing sweeps.}
\label{fig:dist}
\end{figure}
We define the squared distance from the Theoretical Relationship of Eq.~(\ref{topcharge}), $D^{2}_{\textrm{TR}}$, as the minimum distance of each point from the theoretical relationship, i.e. for an instanton candidate with radius $\rho_{0}$ and topological charge at the centre $q_{0}$,
\begin{equation}
D^{2}_{\textrm{TR}} = \min_{\rho}\left[(\rho - \rho_{0})^{2} + (\frac{\pm6}{\pi^{2}\rho^{4}} - q_{0})^{2}\right]\mathrm{.}
\end{equation}
The average value of this is plotted for data from 10 configurations in Fig.~\ref{fig:dist}, as a function of the number of sweeps of overimproved smearing, $\textrm{N}_{\textrm{sw}}$. This confirms our earlier observations; before 50 sweeps, $D^{2}_{\textrm{TR}}$ decreases very rapidly as noise is removed from the lattice. After 50 sweeps, the decrease is characteristically slower.
\par 
\begin{figure}
\includegraphics[trim=1cm 2cm 2cm 2cm, clip=true,height=\hsize,angle=90]{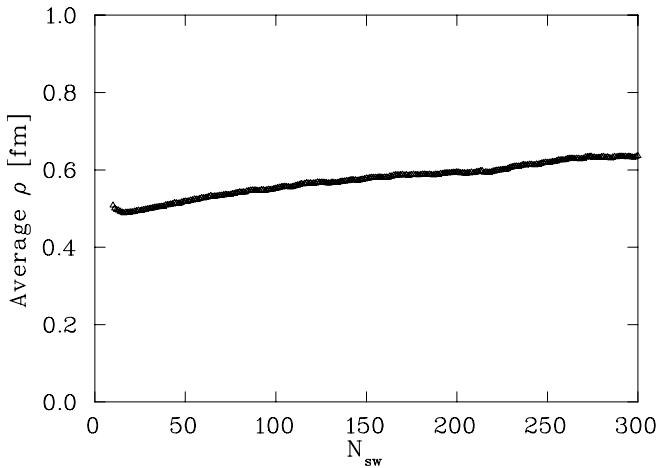}
\caption{Average fitted radius of instanton candidates, as a function of overimproved smearing sweeps.}
\label{fig:radius}
\end{figure}
We have seen in Eq.~(\ref{eq:SLsmac}) that smearing carries the risk of distorting the vacuum structure by enlarging topological objects. This concern is confirmed in Fig.~\ref{fig:radius}, illustrating the average radius of instanton candidates as a function of smearing. After an initial rapid drop, attributable to false positives being rapidly removed by the smearing algorithm, there is a small but steady increase in the average radius of the instanton candidates found.
\par 
\begin{figure}
\includegraphics[trim=1cm 2cm 2cm 2cm, clip=true,height=\hsize,angle=90]{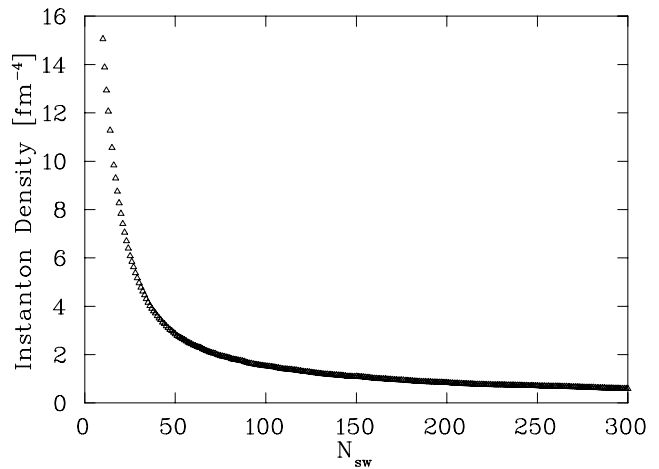}
\caption{Average density of instanton candidates on the lattice, as a function of overimproved smearing sweeps.}
\label{fig:number}
\end{figure}
In Fig.~\ref{fig:number}, we plot the average density of instanton candidates, $N_{inst}/V$, where $N_{inst}$ is the number of instanton candidates found on a configuration and $V$ the lattice volume. This shows a rapid decrease until around 70 sweeps in the early stages of smearing as false positives are rapidly eliminated. Eventually, instanton/anti-instanton annihilation becomes the dominant factor, slowing as the population becomes sparse. Again, the regime of 50 sweeps characterises this transition. Also of note is the  density of $\approx 2\, \mathrm{fm}^{-4}$ at this point. In the instanton liquid model, phenomenological constraints set the instanton radius to be around $\rho=\frac{1}{3} \, \mathrm{fm}$ and density to $N = 1 \, \mathrm{fm}^{-4}$, leading to a packing fraction, the proportion of the vacuum composed of instantons, of $NV_{inst}=N\frac{\pi^{2}\rho^{4}}{2}=0.05$. Here, $V_{inst}=\frac{\pi^{2}\rho^{4}}{2}$ is the 4-volume of an instanton of radius $\rho$. On the lattice, we have found a density of between $0.4 \,\textrm{fm}^{-4}$ and $2 \,\textrm{fm}^{-4}$, reaching an equivalent density to the instanton liquid model at approximately 160 sweeps of smearing. In the second half of this study, we use configurations with less than 100 sweeps of smearing, leading to a density much higher than used in the instanton liquid model.
\par
\begin{figure}
\includegraphics[trim=1cm 2cm 2cm 2cm, clip=true,height=\hsize,angle=90]{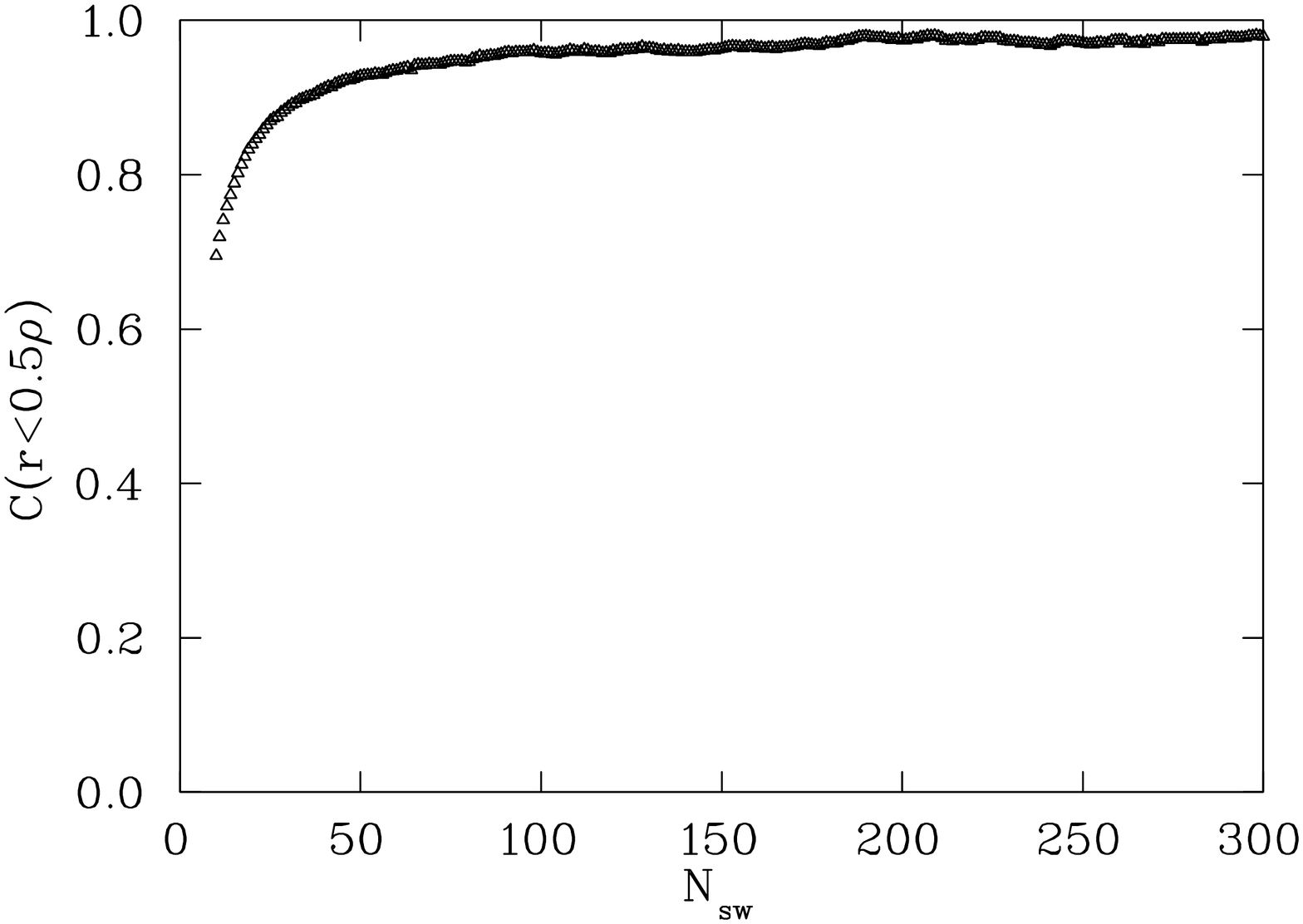}
\includegraphics[trim=1cm 2cm 2cm 2cm, clip=true,height=\hsize,angle=90]{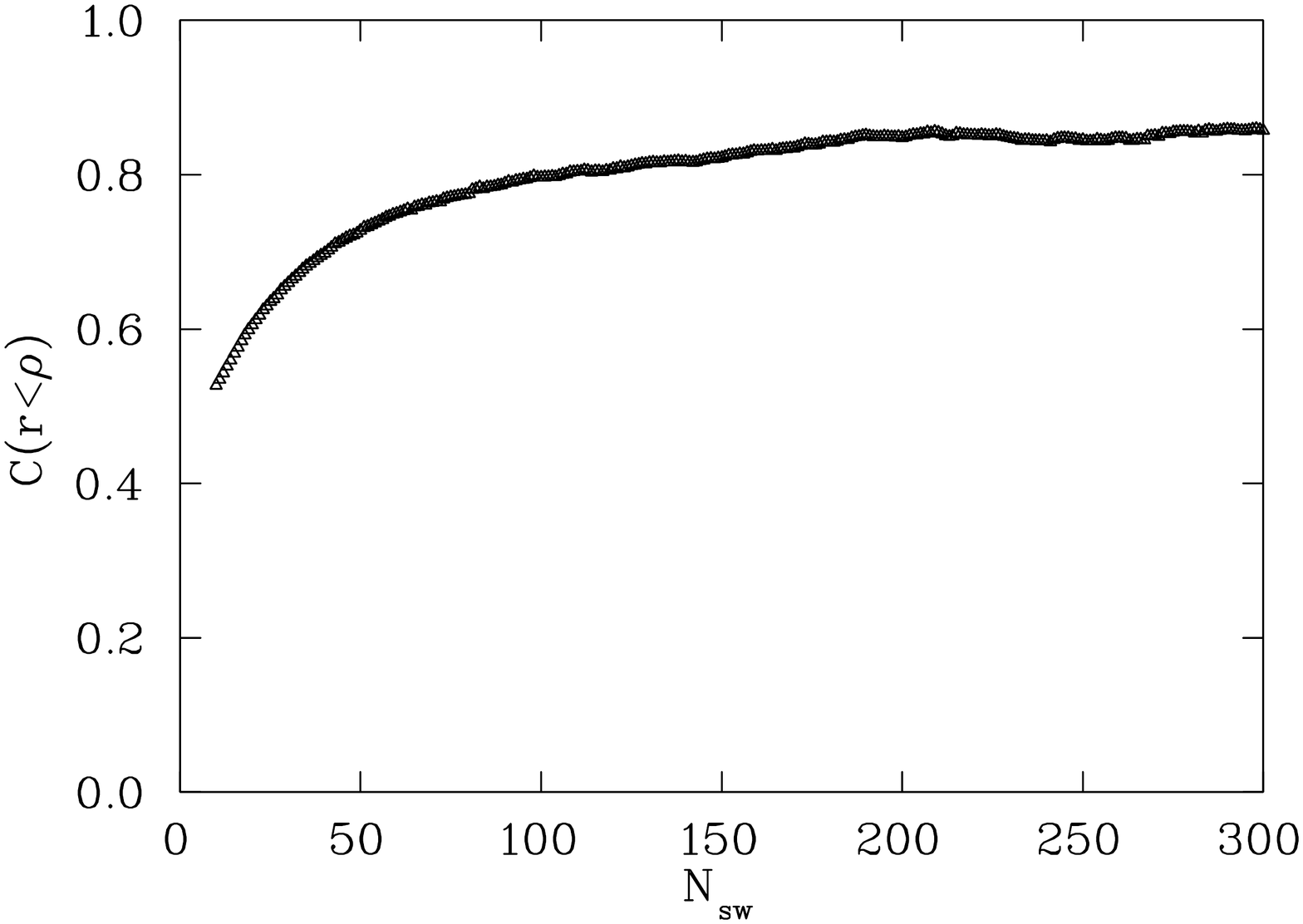}
\caption{The sign coherence, $C$, of topological charge density at the centre of (anti-)instantons is explored as a function of smearing sweeps, $N_{\textrm{sw}}$. Plotted are the fraction of lattice sites within $50\%$ (upper) and $100\%$ (lower) of the fitted radius of an instanton candidates' centre with topological charge of the same sign.}
\label{fig:cohere}
\end{figure}
\begin{figure}
\includegraphics[trim=1cm 2cm 2cm 2cm, clip=true,height=\hsize,angle=90]{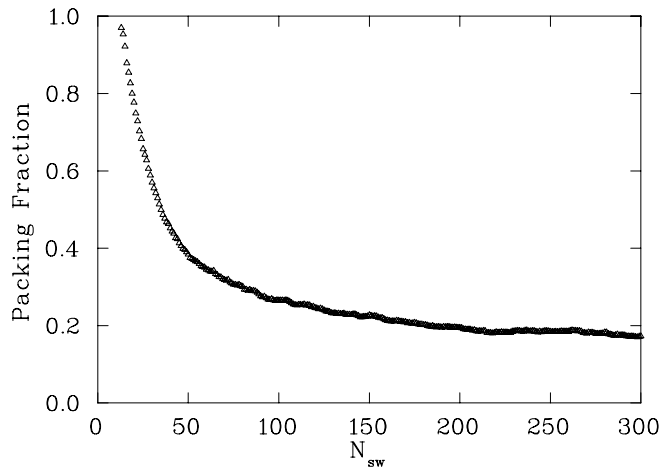}
\caption{Packing fraction of instantons, the percentage of the vacuum composed of instantons, as a function of smearing sweeps.}
\label{fig:packing}
\end{figure}

In order to investigate this further, we have plotted the topological charge coherence, the proportion of lattice sites with topological charge of the same sign as at the centre, within $50\%$, $C(r<0.5\rho)$, and $100\%$, $C(r<\rho)$, of the fitted instanton radius, in Fig.~\ref{fig:cohere}. At $50\%$ of the instanton radius, almost all sites are charge coherent after a relatively small number of smearing sweeps, whereas at $100\%$ of instanton radius, only $85\%$ of the sites are charge coherent, even after a large amount of smearing. This suggests that our fitted values of $\rho$ may be over-estimating the true radius of instantons. We thus calculate the packing fraction using the topological charge found at the centre and Eq.~(\ref{topcharge}). This is plotted in Fig.~\ref{fig:packing}. We note that a packing fraction exceeding 1 is not an issue, as instanton candidates may overlap, particularly at low levels of smearing. We find a packing fraction of between $0.15$ and $0.3$. Once again, this is notably higher than the instanton liquid model value of $0.05$, suggesting that we have found a far "fuller" vacuum on the lattice.
\par
Aggregating these measures, it is clear that between 50 and 100 sweeps of smearing gives an optimal balance, whereby the gauge field is dominated by instantons without excessive distortion. We will thus calculate quark propagators at 0, 30, 50, 80 and 100 sweeps of smearing.

\section{The Overlap Propagator}
\label{sec:overlap}

The overlap fermion operator \cite{Narayanan:1992wx,Narayanan:1993sk,Narayanan:1993ss,Narayanan:1994gw} is defined in the massless case by
\begin{equation}
\label{overlapdef}
D_{o}(0) = \frac{1}{2}(1+\gamma_{5}\epsilon(\gamma_{5}D(m_{w}))),
\end{equation}
with $\epsilon(A) = \frac{A}{\sqrt{A^{2}}}$ the matrix sign function, and the overlap kernel, $D(m_{w})$, any reasonable Hermitian Dirac operator with mass parameter $-m_{w}$, governing the resolution of toplogically non-trivial field structures \cite{Moran:2010rn,Edwards:1998wx}. The overlap is an explicit solution of the Ginsparg-Wilson relation \cite{Ginsparg:1981bj},
\begin{equation}
\label{GWrel}
\gamma_{5}D_{o} + D_{o}\gamma_{5} = 2D_{o}\gamma_{5}D_{o},
\end{equation}
and will thus have a lattice-deformed version of chiral symmetry. This is sufficient to prevent additive quark mass renormalisation \cite{Hasenfratz:1998jp}, greatly simplifying propagator analysis \cite{Skullerud:2001aw,Bonnet:2002ih,Skullerud:2000un}.
\par 
The massive overlap operator is then given by
\begin{equation}
\label{massov}
D_{o}(\mu) = (1-\mu)D_{o}(0) + \mu ,
\end{equation}
where the overlap mass parameter $\mu$ is defined to represent a bare quark mass
\begin{equation}
\label{eq:mudef}
m_{0} = 2m_{w}\mu .
\end{equation}
\par 
We use the FLIC action \cite{Zanotti:2001yb} as the overlap kernel, as studies have shown it to have superior spectral properties, accelerating calculation of the overlap operator, and redcued lattice discretisation errors \cite{Kamleh:2001ff,Kamleh:2004aw}. We project low modes of the kernel and calculate their contribution to the propagator explicitly, greatly reducing the condition number of the matrix square root.\par
\begin{figure}
\includegraphics[trim=1cm 2cm 2cm 1.5cm, clip=true,height=\hsize,angle=90]{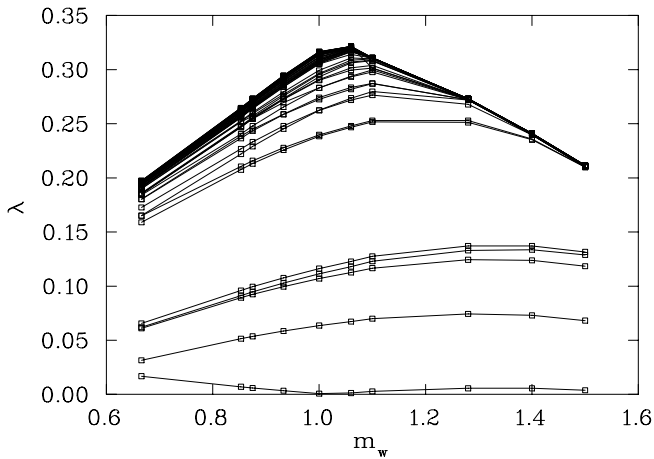}
\caption{The 50 lowest-lying eigenvalues of the overlap kernel as a function of kernel mass parameter $m_{w}$, from a single configuration with fifty sweeps of smearing.}
\label{eigenflow}
\end{figure}

Notably, we have found results consistent with the work of Neuberger \cite{Neuberger:1999pz}, who predicts that in the presence of a sufficiently smooth background field, the eigenvalues of the Wilson kernel are maximally displaced from $0$ at $m_{w}=1$. In Fig.~\ref{eigenflow}, we have plotted the 50 lowest lying eigenvalues of the FLIC kernel on a configuration with 50 sweeps of smearing and we observe that the region where the spectrum becomes dense(i.e. ignoring the isolated low-lying topological nodes) is indeed maximally separated from 0 near $m_{w}=1$. We choose to perform all calculations at $m_{w}=1$.\par
We consider nine values of $\mu$, given in Table \ref{mutable}.
\par
\begin{table}
\begin{center}
\begin{ruledtabular}
\begin{tabular}[c]{cc}
$\mu$ & $m_{bare}$(MeV) \\ \hline
0.01271 & 39.8\\ 
0.01694 & 53.0\\ 
0.02119 & 66.4\\ 
0.02543 & 80.0\\ 
0.02966 & 93.0\\ 
0.03390 & 106.2\\ 
0.04238 & 132.7\\ 
0.05076 & 159.0\\ 
0.06356 & 199.1\\
\end{tabular}
\end{ruledtabular}
\end{center}
\caption{The nine values of $\mu$ considered, with the corresponding bare mass in physical units}
\label{mutable}
\end{table}

In order to define an overlap quark propagator, we naturally wish to preserve the most important properties of the continuum propagator. As a simple consequence of chiral symmetry, in the massless case the continuum quark propagator obeys
\begin{equation}
\{\gamma_5,S^{\textrm{cont}}_{m=0}(p)\} = 0.
\end{equation}
We also note that Eq.~(\ref{GWrel}) implies
\begin{equation}
\{\gamma_{5},D_{o}^{-1}\} = 2\gamma_{5},
\end{equation}
suggesting an appropriate form for the overlap propagator,
\begin{equation}
S(p)_{m=0} \propto (D_{o}^{-1} - 1).
\end{equation}
It is also important to maintain the correct continuum limit, and so given
\begin{equation}
\lim_{a \rightarrow 0} D_{o} = \frac{1}{2m_{w}}D\!\!\!\!/\ ,
\end{equation}
it is then natural to define the massless bare overlap propagator as \cite{Narayanan:1994gw,Edwards:1998wx}
\begin{equation}
\label{eq:baremassoverlap}
S(p)_{m=0} = \frac{1}{2m_{w}}(D_{o}^{-1} - 1).
\end{equation}
We can then construct the bare massive overlap propagator simply by adding a bare mass to the inverse of Eq.~(\ref{eq:baremassoverlap}),
\begin{equation}
S(p)_{m=m_{0}}^{-1} = S(p)_{m=0}^{-1} + m_{0},
\end{equation}
then recalling the definition of the overlap mass parameter in Eq.~(\ref{eq:mudef}), we have
\begin{equation}
\label{massoverprop}
S(p) = \frac{1}{2m_{w}(1-\mu)}(D_{o}^{-1} - 1).
\end{equation}
Due to the lack of additive mass renormalisation, the quark propagator on the lattice will have as its general form
\begin{equation}
\label{propgenform}
S(p) = \frac{Z(p)}{iq\!\!\!/\ + M(p)},
\end{equation}
with $M(p)$ the non-perturbative mass function and $Z(p)$ containing all renormalisation information. \par 
This can be defined as having the inverse
\begin{equation}
\label{propingenform}
S^{-1}(p) \equiv i\sum_{\mu}(C_{\mu}(p)\gamma_{\mu}) + B(p) \equiv iq\!\!\!/\ A(p) + B(p),
\end{equation}
where $q\!\!\!/\ $ is defined\cite{Bonnet:2002ih} by the tree level propagator, calculated with $U_{\mu}(x)=1$,
\begin{equation}
S_{0}^{-1}(p) = iq\!\!\!/\ + m_{0}.
\end{equation}
Comparing Eq.~(\ref{propingenform}) to Eq.~(\ref{propgenform}), we see
\begin{eqnarray}
Z(p) &=& \frac{1}{A(p)} \nonumber \\
M(p) &=& \frac{B(p)}{A(p)}.
\end{eqnarray}
Multiplying Eq.~(\ref{propingenform}) by $q\!\!\!/\ $ and taking the trace provides
\begin{equation}
A(p) = \frac{q_{\mu}C_{\mu}}{q^{2}}.
\end{equation}
To determine $A(p)$ and $B(p)$ we follow Ref.~\cite{Kamleh:2004aw} and define $\mathcal{C}$ and $\mathcal{B}$ by
\begin{equation}
S(p) = -i\mathcal{C}\!\!\!\!/\ (p) + \mathcal{B}(p),
\end{equation}
such that
\begin{eqnarray}
\mathcal{B}(p) &=& \frac{1}{n_{s}n_{c}}\mathrm{Tr}(S(p)) \nonumber \\
\mathcal{C}_{\mu}(p) &=& \frac{i}{n_{s}n_{c}}\mathrm{Tr}[\gamma_{\mu}S(p)],
\end{eqnarray}
where $n_{s}$ and $n_{c}$ are the extent of spin and colour indices. \par 
Comparing to Eq.~(\ref{propingenform}), we find
\begin{eqnarray}
C_{\mu}(p) &=& \frac{\mathcal{C}_{\mu}(p)}{\mathcal{C}^{2}(p) + \mathcal{B}^{2}(p)} \nonumber \\
B(p) &=& \frac{\mathcal{B}(p)}{\mathcal{C}^{2}(p) + \mathcal{B}^{2}(p)}.
\end{eqnarray}
Defining
\begin{equation}
\mathcal{A}(p) = \frac{q_{\mu}\mathcal{C}_{\mu}}{q^{2}},
\end{equation}
we thus have
\begin{eqnarray}
Z(p) &=& \frac{\mathcal{C}^{2}(p) + \mathcal{B}^{2}(p)}{\mathcal{A}(p)} \nonumber \\
M(p) &=& \frac{\mathcal{B}(p)}{\mathcal{A}(p)}.
\end{eqnarray}

\section{Results}
\label{sec:results}

We fix to Landau gauge by maximizing the $\mathcal{O}(a^{2})$ improved gauge fixing functional\cite{Davies:1988pr}
\begin{equation}
\mathcal{F}_{\textrm{imp}} = \sum_{x,\mu}\textrm{Re}\textrm{Tr}{\big(}\frac{4}{3}U_{\mu}-\frac{1}{12u_{0}}(U_{\mu}(x)U(x+\hat{u})+\textrm{H.C.}){\big)}
\end{equation}
using a Fourier transform accelerated algorithm\cite{Roberts:2010cz,Roberts:2012em,Bonnet:1999mj,Davies:1988pr}. To avoid Gribov copy issues, we first gauge fix configurations with 100 sweeps of smearing and then use these as a preconditioner for the same configurations with lower levels of smearing\cite{Hetrick:1997yy}.
The matrix sign function is calculated using the Zolotarev rational polynomial approximation \cite{Chiu:2002hh}. We average data over spatial symmetries and choose $p=5\,\mathrm{GeV}$ as the renormalization point for $Z^{(\mathcal{R})}$, with $Z^{(\mathcal{R})}(p=5\textrm{Gev})=1$. We apply a cylinder cut \cite{Bonnet:2002ih} to data with a radius of $\frac{\pi}{2a}$, and estimate errors using a second-order single-elimination jack-knife method. \par 
\begin{figure}
\includegraphics[trim=1cm 2cm 2cm 2.5cm, clip=true,height=\hsize,angle=90]{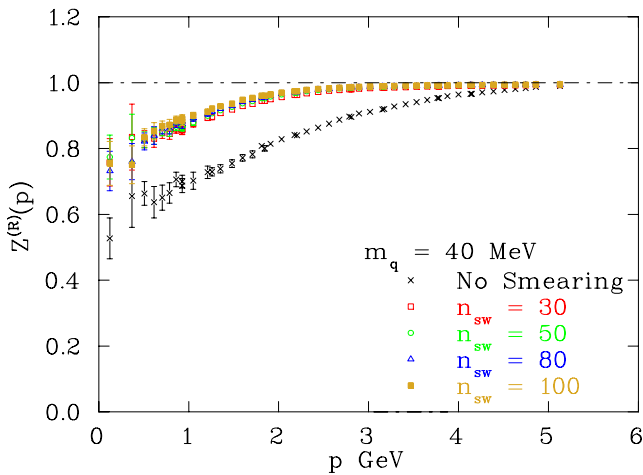}
\caption{The renormalization function at $m_{q} = 39.8 \, \textrm{MeV}$, on configurations with 0, 30, 50, 80, and 100 sweeps of smearing.}
\label{Z1271}
\end{figure}
$Z^{(\mathcal{R})}$ is plotted in Fig.~\ref{Z1271} for the lightest quark mass, $39.8 \, \mathrm{MeV}$, for various levels of smearing. This reveals the well-known shape of the renormalization function for the unsmeared case; dipping in the infrared region and rising in the ultraviolet limit. After just 30 sweeps of smearing however, different behaviour is seen. The renormalisation function is more tree like, dipping half the amount and approximating $1$ for $p > 3 \, \textrm{GeV}$. The renormalisation function rises earlier to this plateau value, with similar behaviour seen at all levels of smearing. This appears to confirm our earlier observations; after 30 sweeps most short-range behaviour is removed, and the gauge field structure of the lattice is much simpler. The physics responsible for the drop in $Z^{(\mathcal{R})}$ is removed early as all curves agree. \par
\begin{figure}
\includegraphics[trim=1cm 2cm 2cm 2.5cm, clip=true,height=\hsize,angle=90]{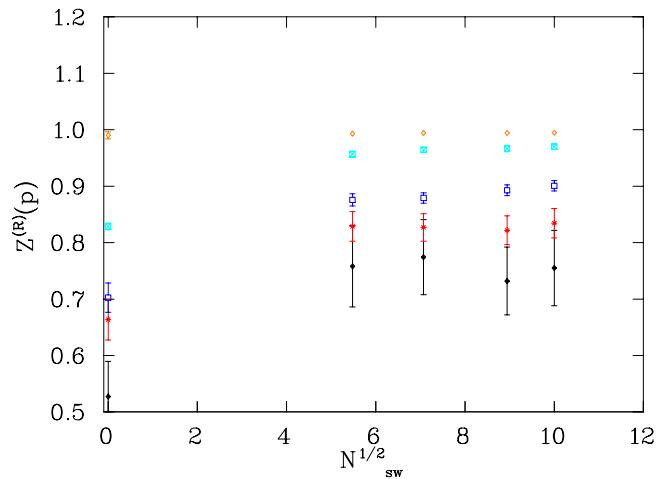}
\caption{The renormalization function plotted as a function of $N_{sw}^{1/2}$ at $p$ values of 0.3, 0.6, 1, 2 and 5 GeV (ascending order), with $m_{q} = 40 \, \textrm{MeV}$.}
\label{Z1271indq}
\end{figure}
To further show this, $Z^{(\mathcal{R})}$ is plotted as a function of $\sqrt{N_{sw}}$ at given values of $p$ in Fig.~\ref{Z1271indq}. We have chosen to plot against $\sqrt{N_{sw}}$ as this is proportional to the smearing radius. This shows a large change from 0 to 30 sweeps of smearing, affirming that the physics responsible for the drop in $Z^{(\mathcal{R})}$ is removed early in the smearing process, after which smearing has relatively little effect. \par
\begin{figure*} [h!!]
\includegraphics[trim=1cm 2cm 2cm 2.5cm, clip=true,scale=0.26,angle=90]{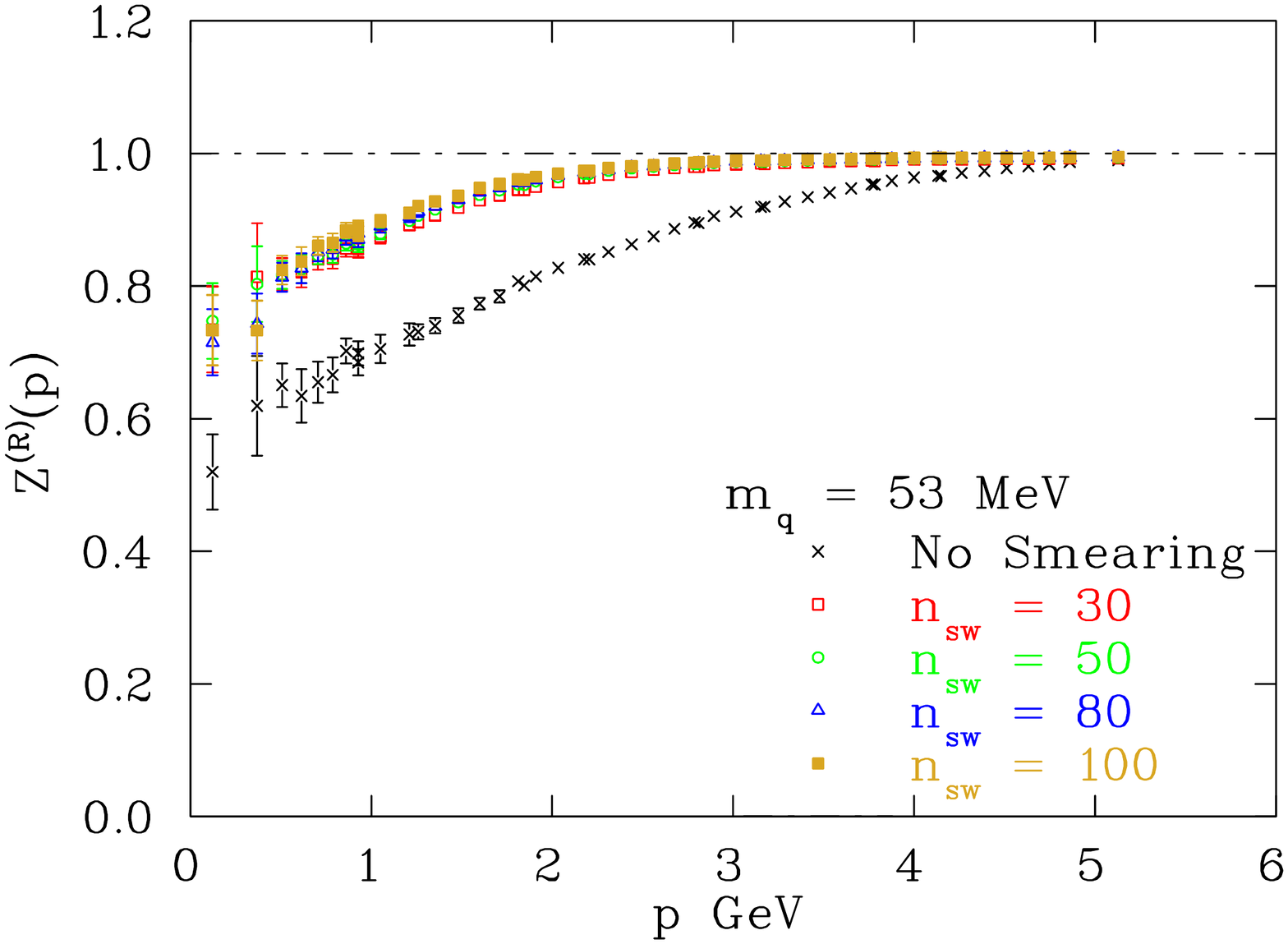}
\includegraphics[trim=1cm 2cm 2cm 2.5cm, clip=true,scale=0.26,angle=90]{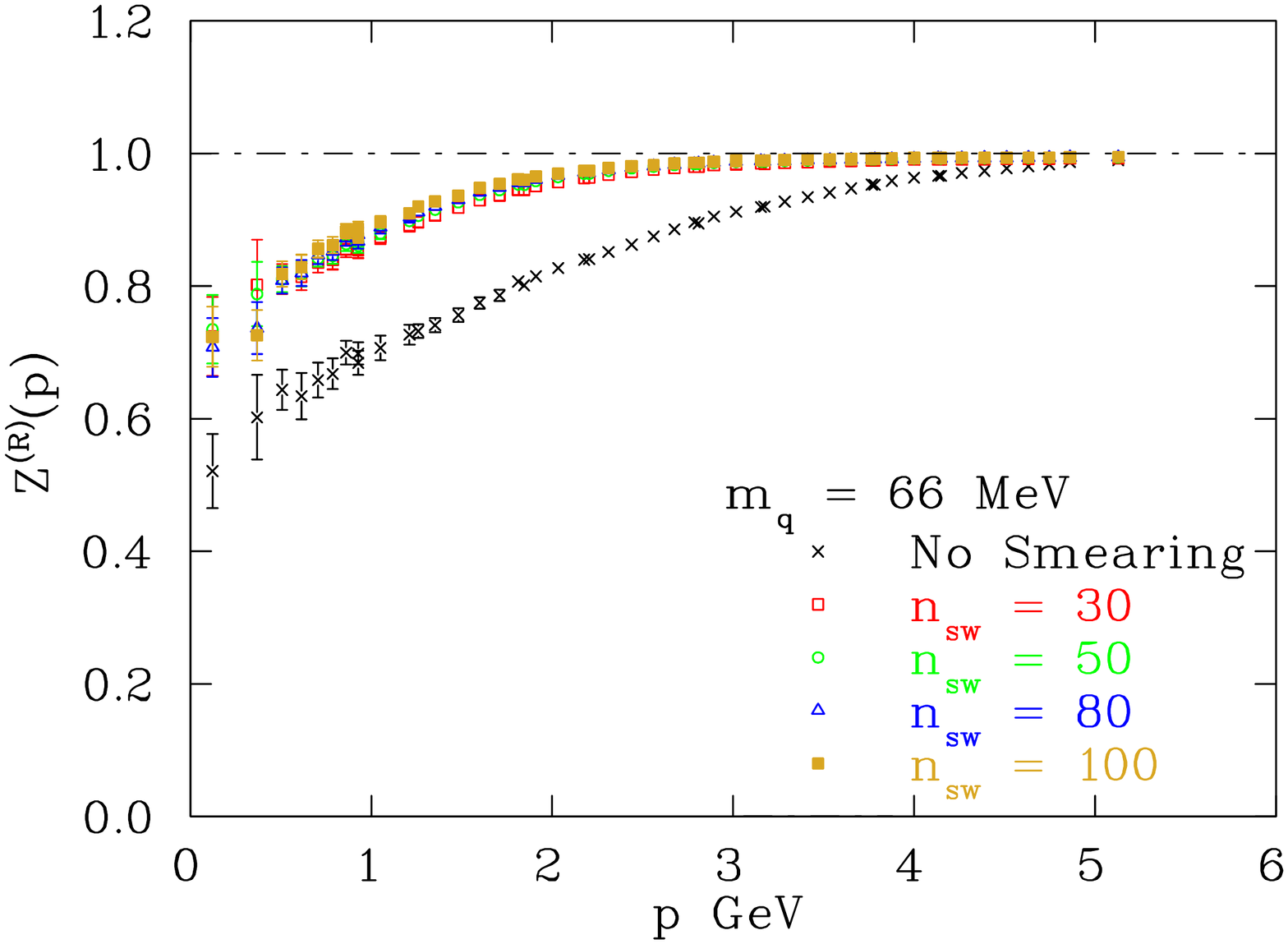}
\includegraphics[trim=1cm 2cm 2cm 2.5cm, clip=true,scale=0.26,angle=90]{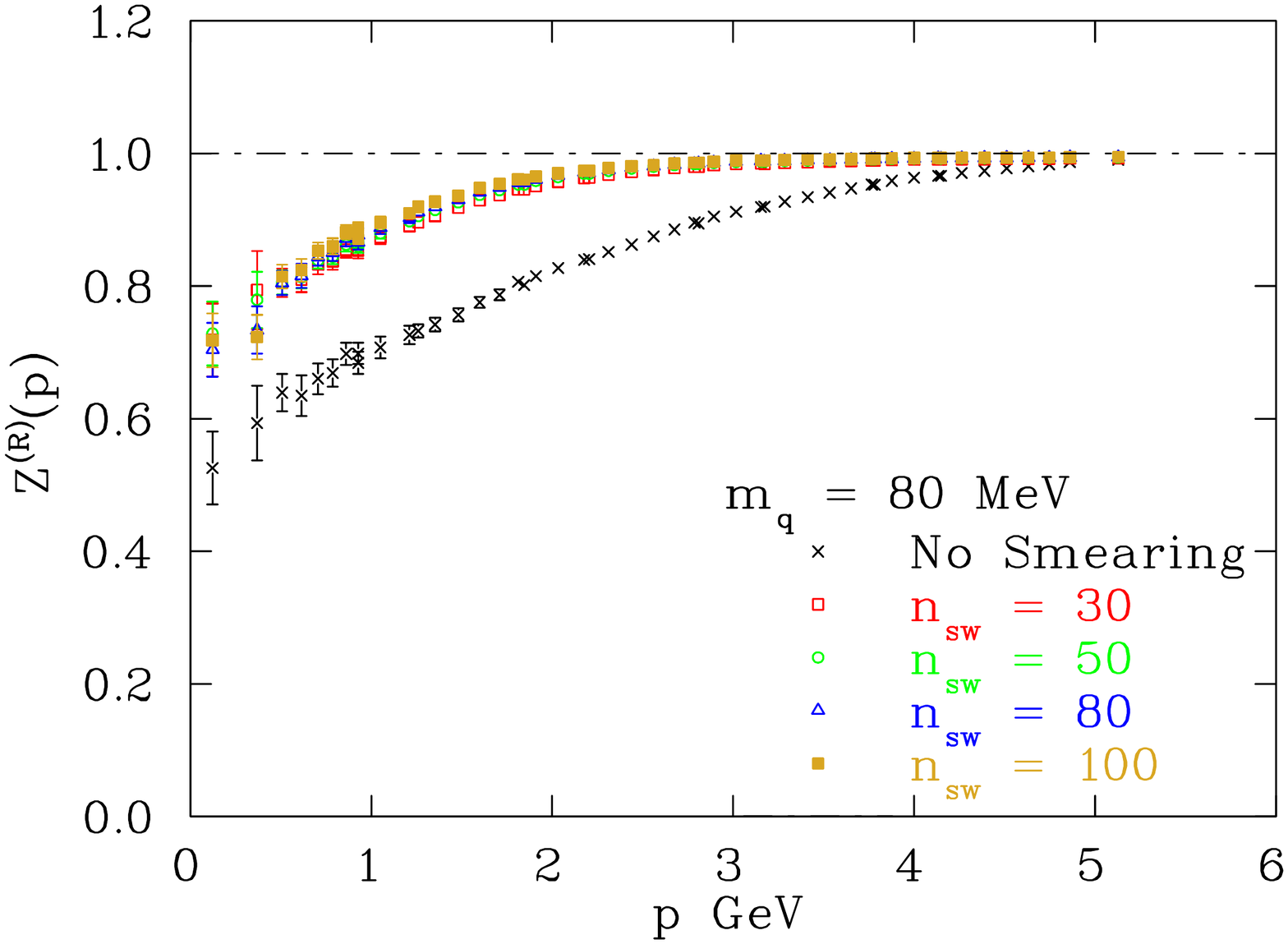}
\includegraphics[trim=1cm 2cm 2cm 2.5cm, clip=true,scale=0.26,angle=90]{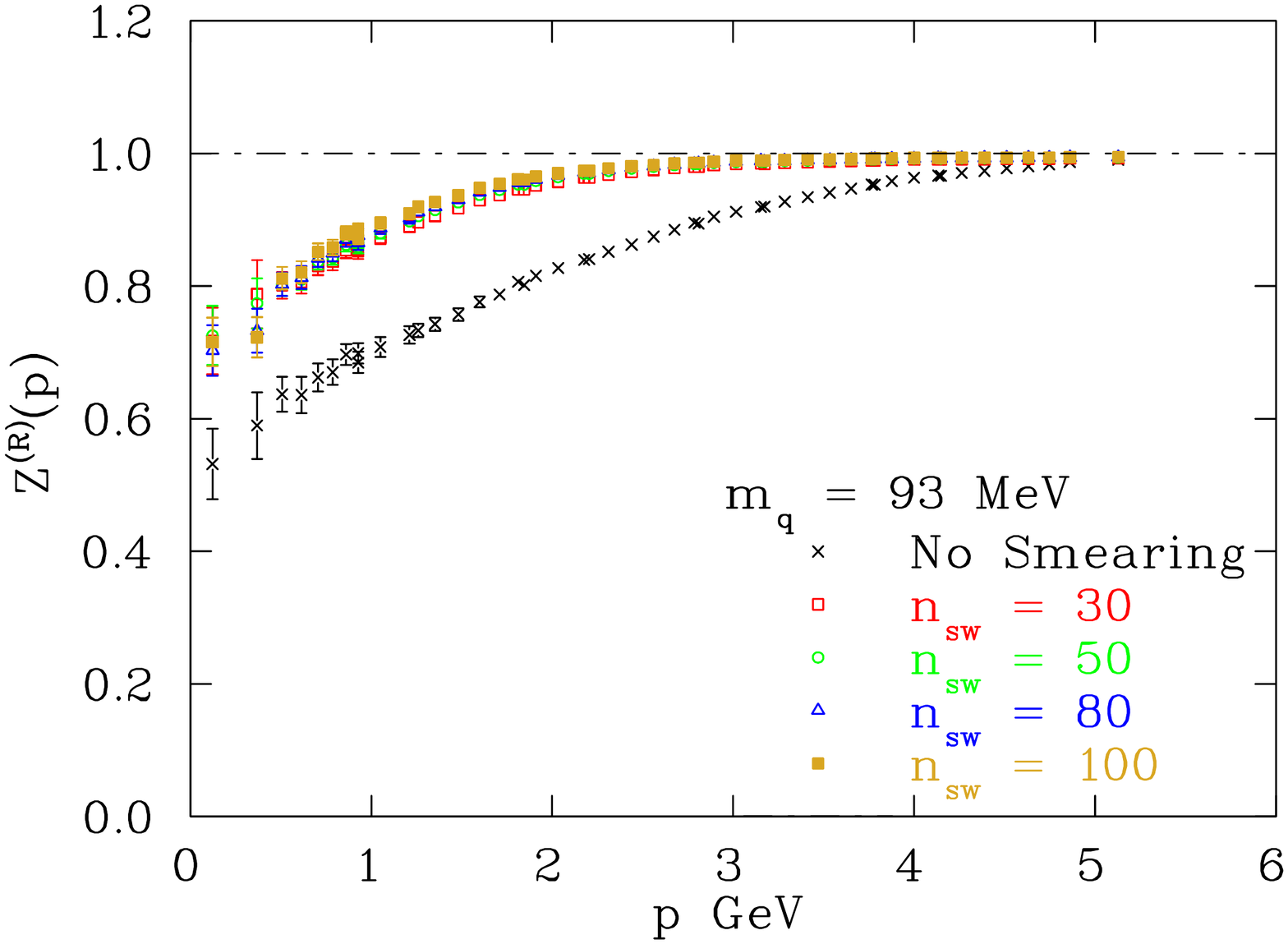}
\includegraphics[trim=1cm 2cm 2cm 2.5cm, clip=true,scale=0.26,angle=90]{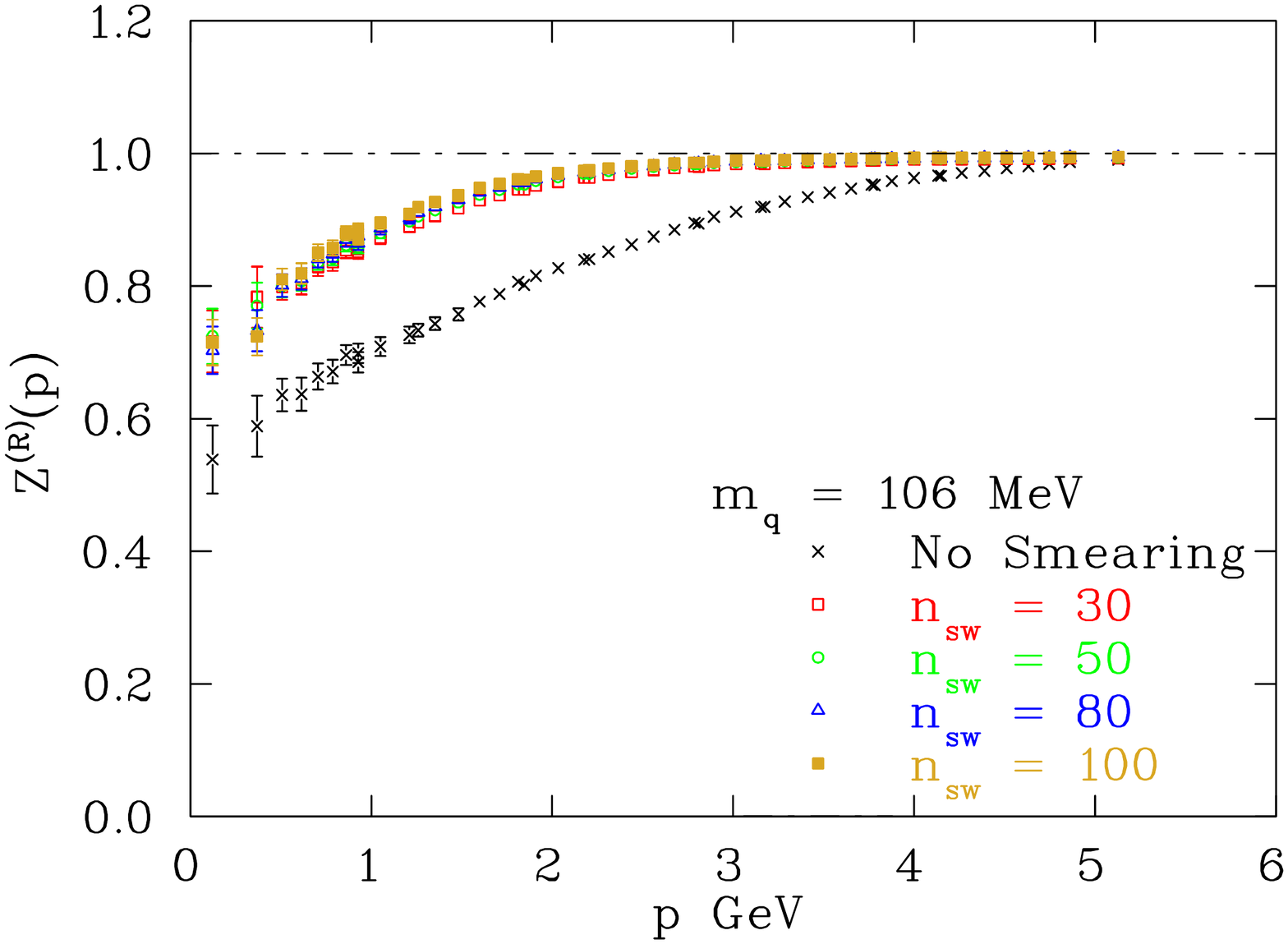}
\includegraphics[trim=1cm 2cm 2cm 2.5cm, clip=true,scale=0.26,angle=90]{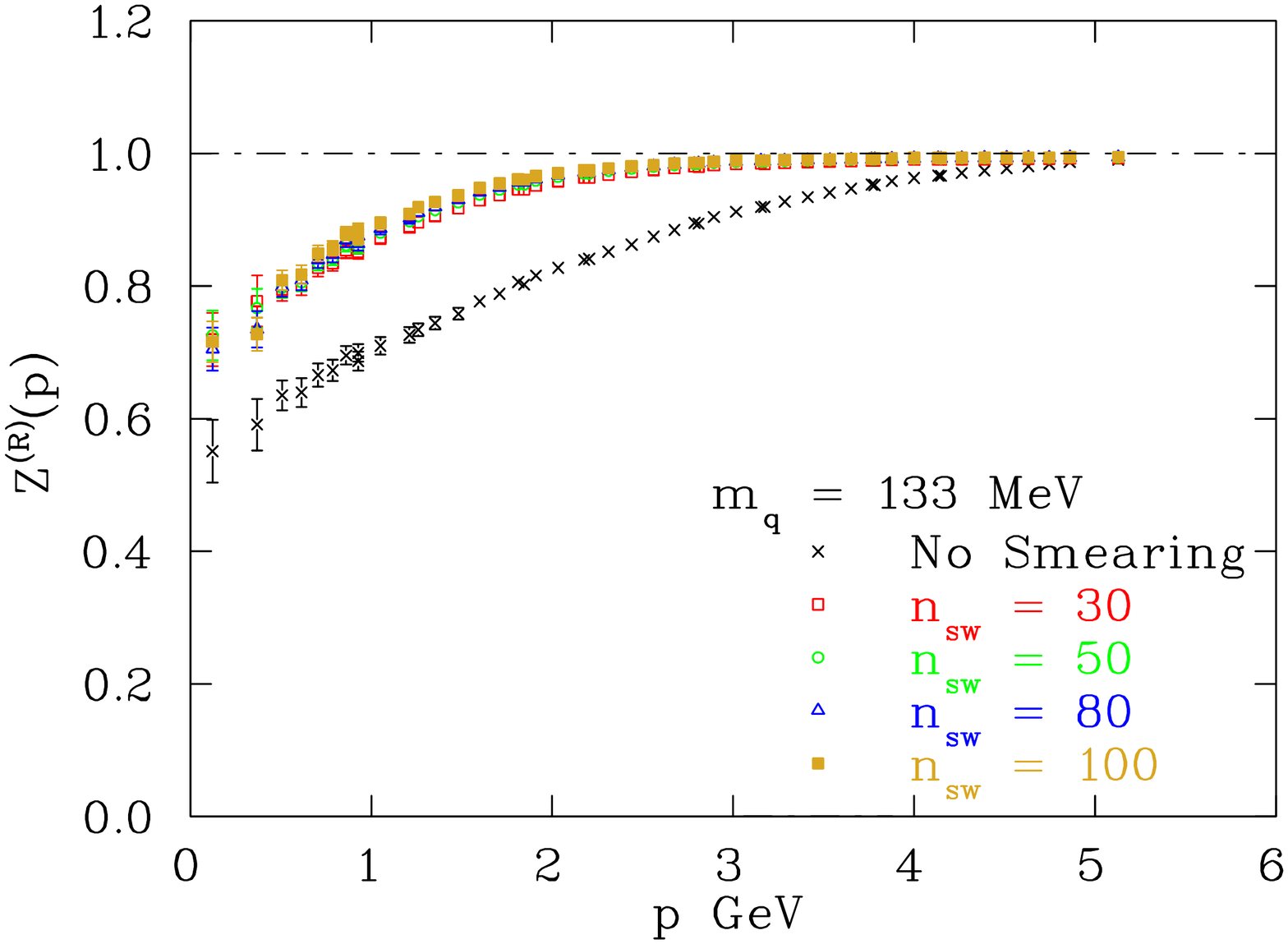}
\includegraphics[trim=1cm 2cm 2cm 2.5cm, clip=true,scale=0.26,angle=90]{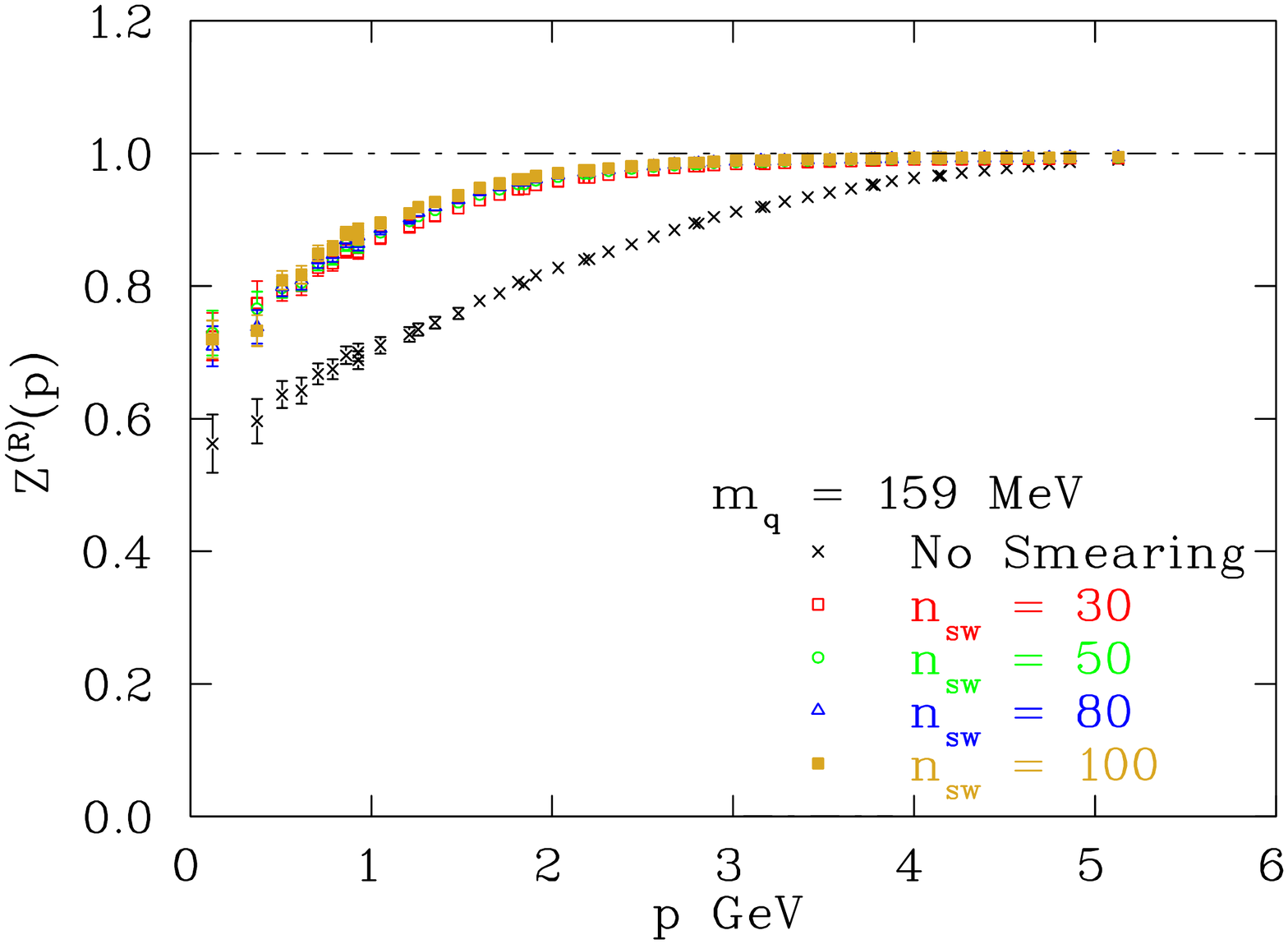}
\includegraphics[trim=1cm 2cm 2cm 2.5cm, clip=true,scale=0.26,angle=90]{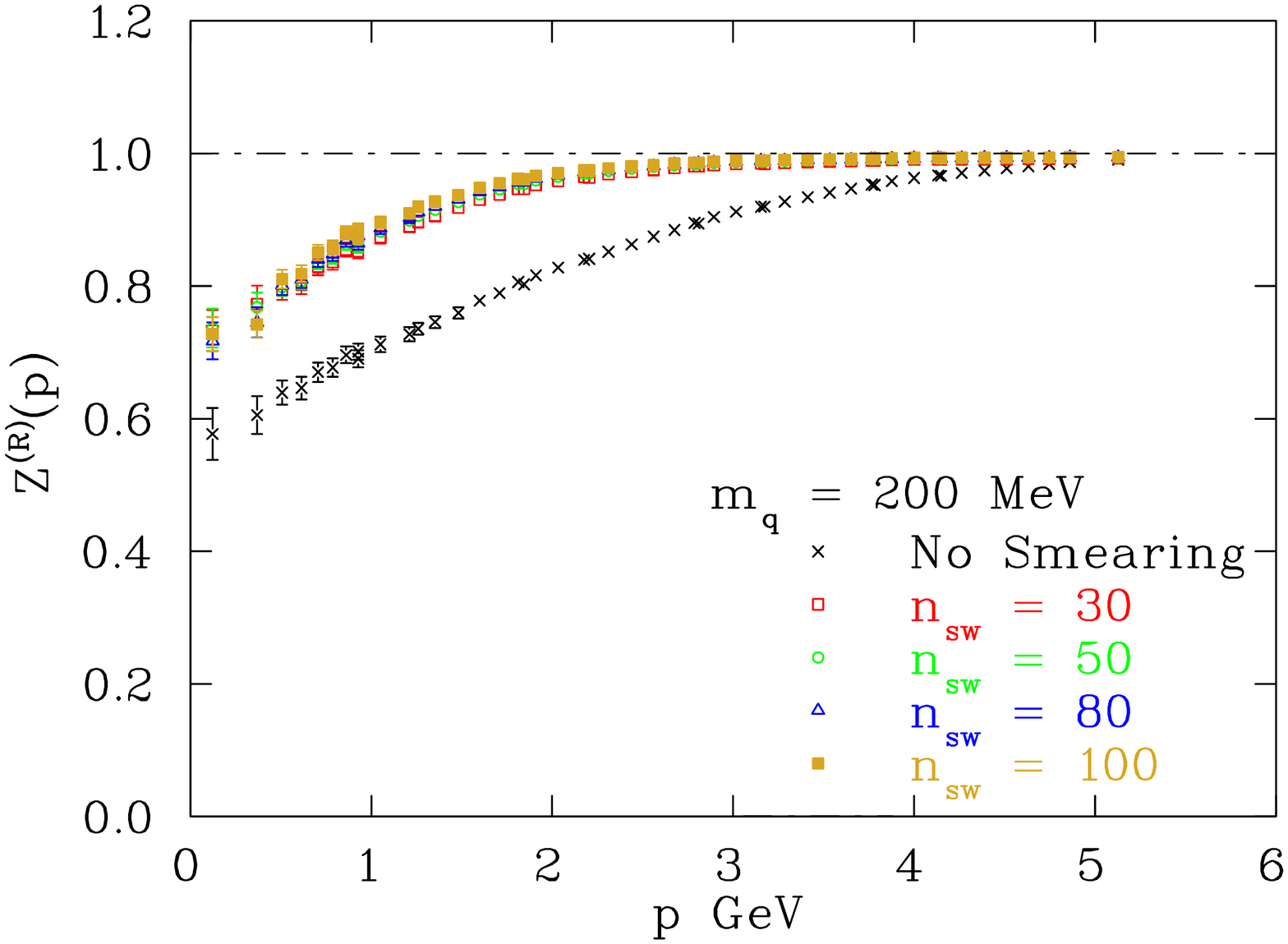}
\caption{The renormalisation function at various values of $m_{q}$, on configurations with 0, 30, 50, 80, and 100 sweeps of smearing.}
\label{fig:Zhigh}
\end{figure*}

Higher masses are illustrated for all $p$ in Fig.~\ref{fig:Zhigh}. Remarkably little sensitivity to the quark mass is observed. We note that the lowest two $p$ values are purely timelike, and thus have comparatively large error bars due to the lack of symmetries for averaging.
\par
\begin{figure}
\includegraphics[trim=1cm 2cm 2cm 2.5cm, clip=true,height=\hsize,angle=90]{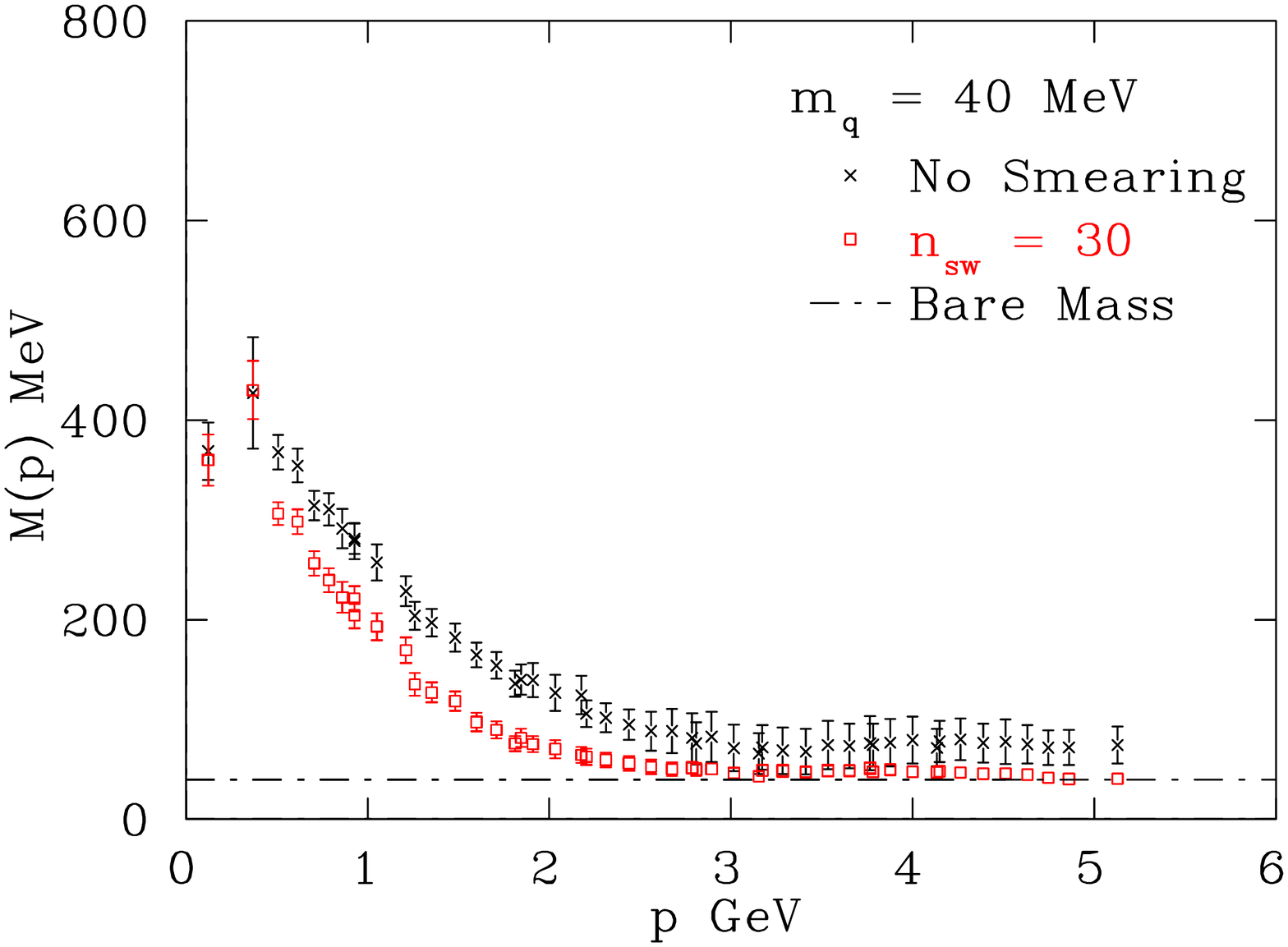}
\caption{The mass function at $m_{q}=39.8 \,\textrm{MeV}$ on unsmeared configurations and with 30 sweeps of smearing.}
\label{M1271thirty}
\end{figure}
We now plot the mass function for a quark mass of $39.8 \, \mathrm{MeV}$ in Fig.~\ref{M1271thirty}. The unsmeared data reveals the expected shape; a large effective quark mass in the infrared region created through dynamical mass generation, tapering to an approximate plateau in the ultraviolet region, where logarithmic corrections produce a running quark mass higher than the input bare mass, illustrated by the dot-dash line.\par 
Looking at the lowest level of smearing considered, 30 sweeps, we note that the smeared results consistently have smaller error bars, particularly in the ultra-violet region, indicating short distance physics is a significant source of noise. Although the instanton content of each ensemble is different, its impact on the mass function remains similar. We also note that, as expected, in the ultraviolet region the running quark mass is now barely higher than the input bare mass, due to the spoiling of short distance perturbative physics at this range. At the lowest momenta, we find perfect agreement with the unsmeared case, but as we increase momentum there is some suppression of dynamically generated mass. We attribute this to the destruction of some topological objects by the smearing algorithm, both those smaller than the dislocation threshold and those eliminated by pair annihilation. Smearing also has the effect of enlarging objects, again removing dynamically generated mass from this region. Indeed one anticipates a significant smearing dependence in the regime of $p \approx 1 \textrm{GeV}$.\par
\begin{figure}
\includegraphics[trim=1cm 2cm 2cm 2.5cm, clip=true,height=\hsize,angle=90]{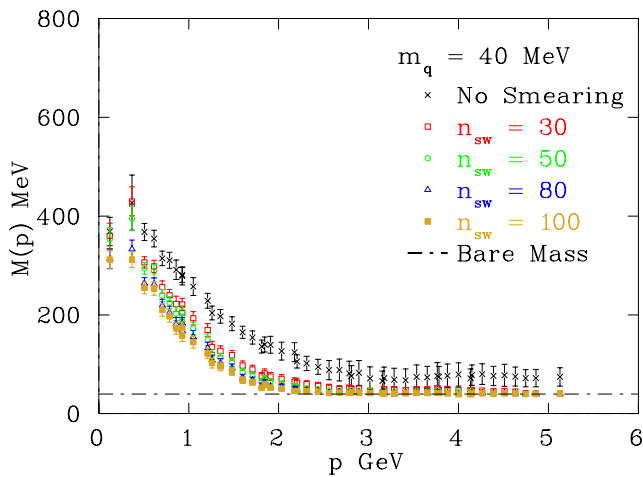}
\caption{The mass function at $m_{q}=39.8 \,\textrm{MeV}$ on unsmeared configurations and with 30, 50, 80 and 100 sweeps of smearing.}
\label{M1271hundred}
\end{figure}
We now plot all levels of smearing, in Fig.~\ref{M1271hundred}. Even at high levels of smearing, we have maintained the qualitative shape of the mass function, although increasing smearing results in a larger loss of dynamically generated mass. As the level of smearing increases, the mass function becomes flat for lower values of momenta. This is in accord with the increase of the smearing radius, the distance within which physics is suppressed.  The most dramatic shift is between the unsmeared case and 30 sweeps. This suggests that, to at least some extent, smearing has changed the mass renormalization, and so in order to acquire results with the same physical mass we must shift our value of $\mu$ for the smeared results.\par
\begin{figure}
\includegraphics[trim=1cm 2cm 2cm 2.5cm, clip=true,height=\hsize,angle=90]{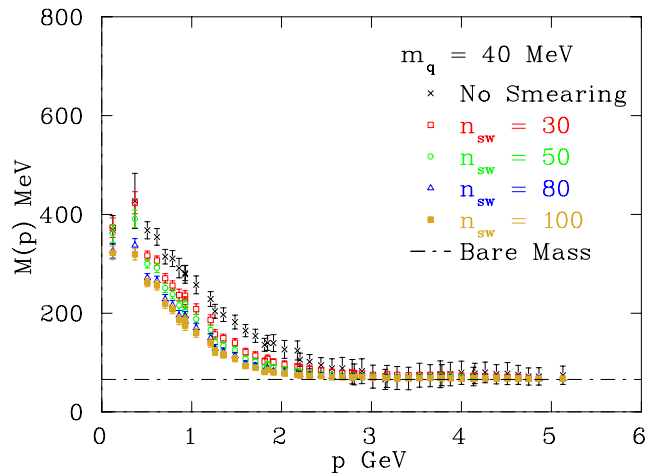}
\caption{The mass function at $\mu = 0.01271$ on unsmeared configurations and at $\mu = 0.02119$ on configurations with 30, 50, 80 and 100 sweeps of smearing.}
\label{Mshifted}
\end{figure}
In Fig.~\ref{Mshifted} we have plotted smeared results at a value of $\mu=0.02119$, chosen in order to match unsmeared results in the ultraviolet limit. This reveals that not all loss of dynamical mass generation is due to a changing mass renormalization. Although we have retained the majority of long-range physics, it is clear that smearing, particularly after a large number of sweeps, has removed some important aspects of vacuum structure, creating a gap between unsmeared and smeared results. \par
\begin{flushleft}
\begin{figure}
\includegraphics[trim=1cm 2cm 2cm 2.5cm, clip=true,height=\hsize,angle=90]{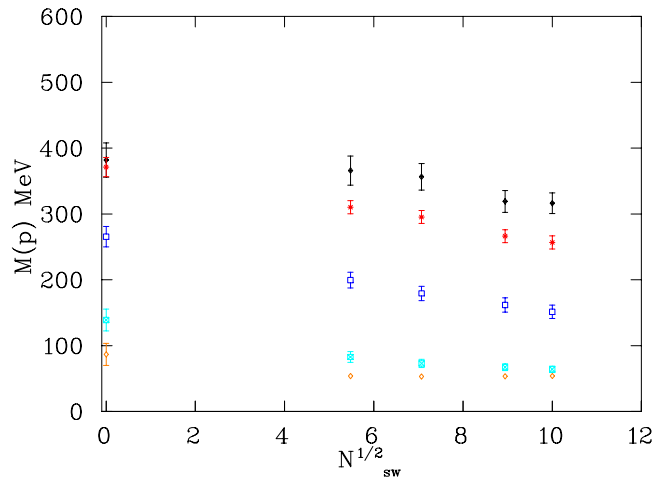}
\caption{The mass function plotted as a function of $N_{sw}^{1/2}$ at $p$ values of 0.3, 0.6, 1, 2 and 5 GeV (descending order), with $m_{q} = 53 \, \textrm{MeV}$.}
\label{M1694indq}
\end{figure}
\end{flushleft}
$M(p)$ is plotted as a function of $\sqrt{N_{sw}}$ at fixed values of $p$ in Fig.~\ref{M1694indq}, for a quark mass of $m_{q} = 53 \, \textrm{MeV}$. This shows an approximately linear dependence on $\sqrt{N_{sw}}$ and thus the smearing radius.\par
This suggests that the increase in smearing radius is responsible for loss of dynamical mass generation by removing topological objects from the lattice. At higher momenta corresponding to smaller distances, smearing has removed all relevant physics from the lattice early, and so an increase in the smearing radius has a smaller effect.\par
\begin{figure*}
\includegraphics[trim=1cm 2cm 2cm 2.5cm, clip=true,scale=0.26,angle=90]{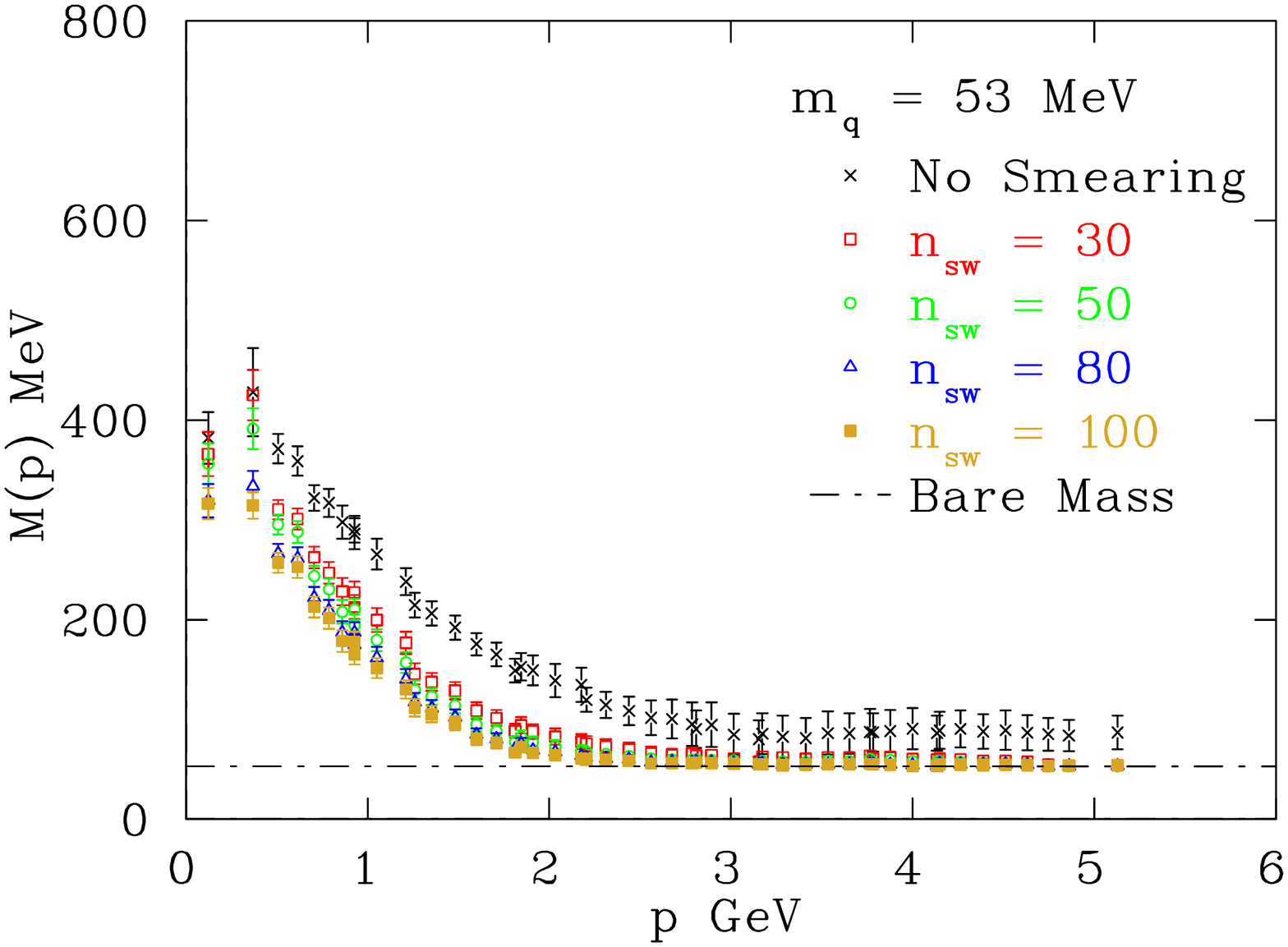}
\includegraphics[trim=1cm 2cm 2cm 2.5cm, clip=true,scale=0.26,angle=90]{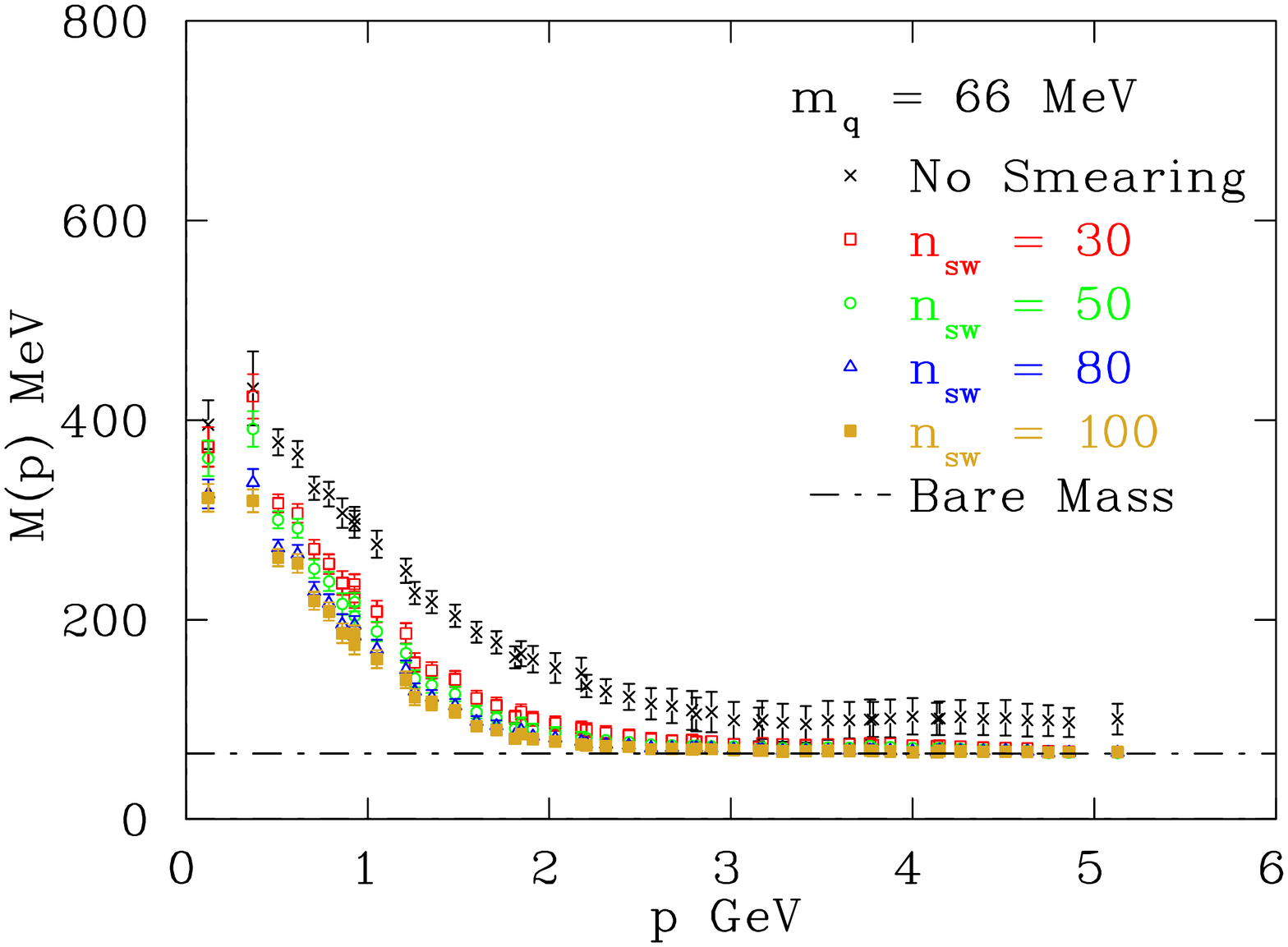}
\includegraphics[trim=1cm 2cm 2cm 2.5cm, clip=true,scale=0.26,angle=90]{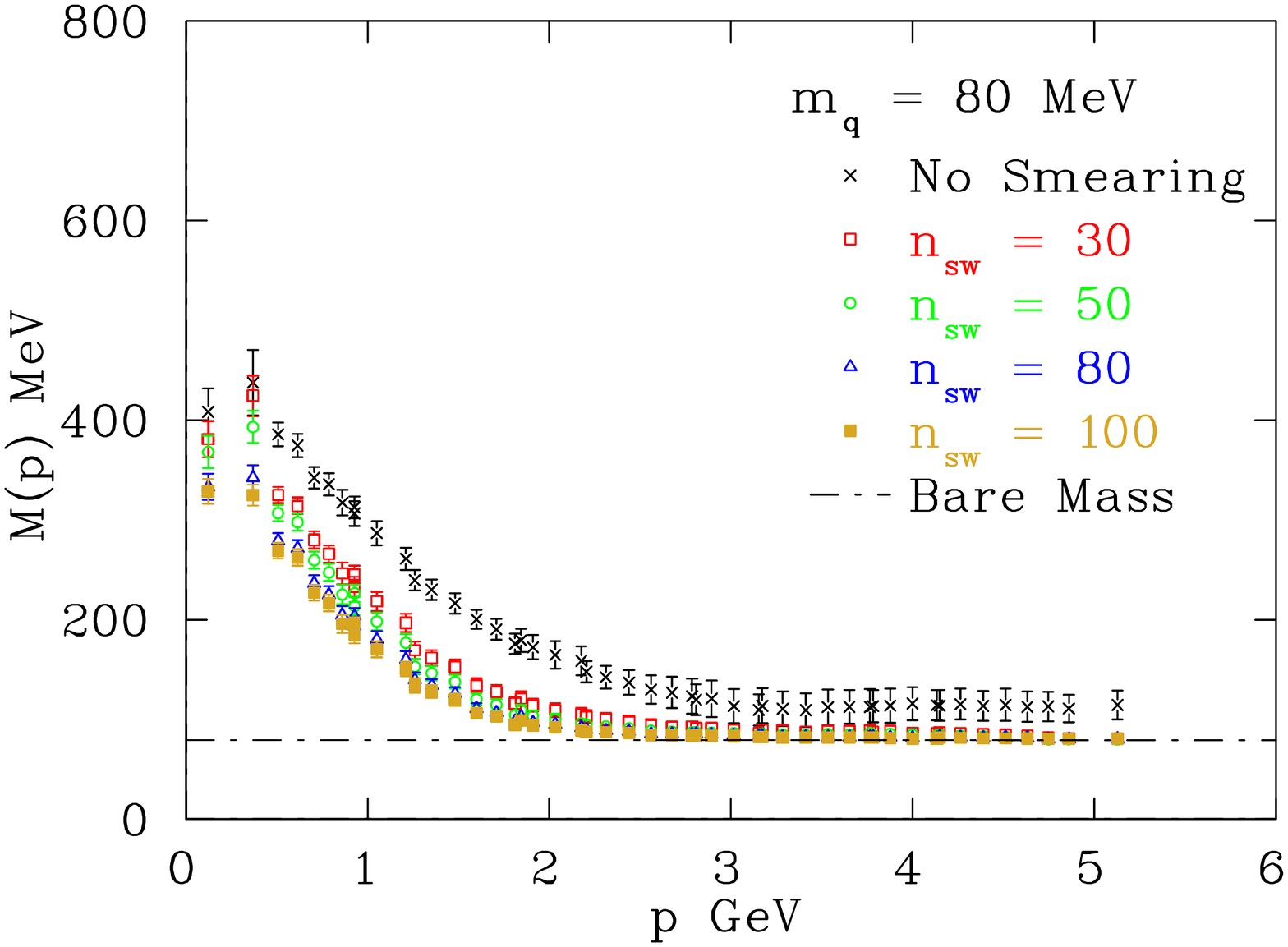}
\includegraphics[trim=1cm 2cm 2cm 2.5cm, clip=true,scale=0.26,angle=90]{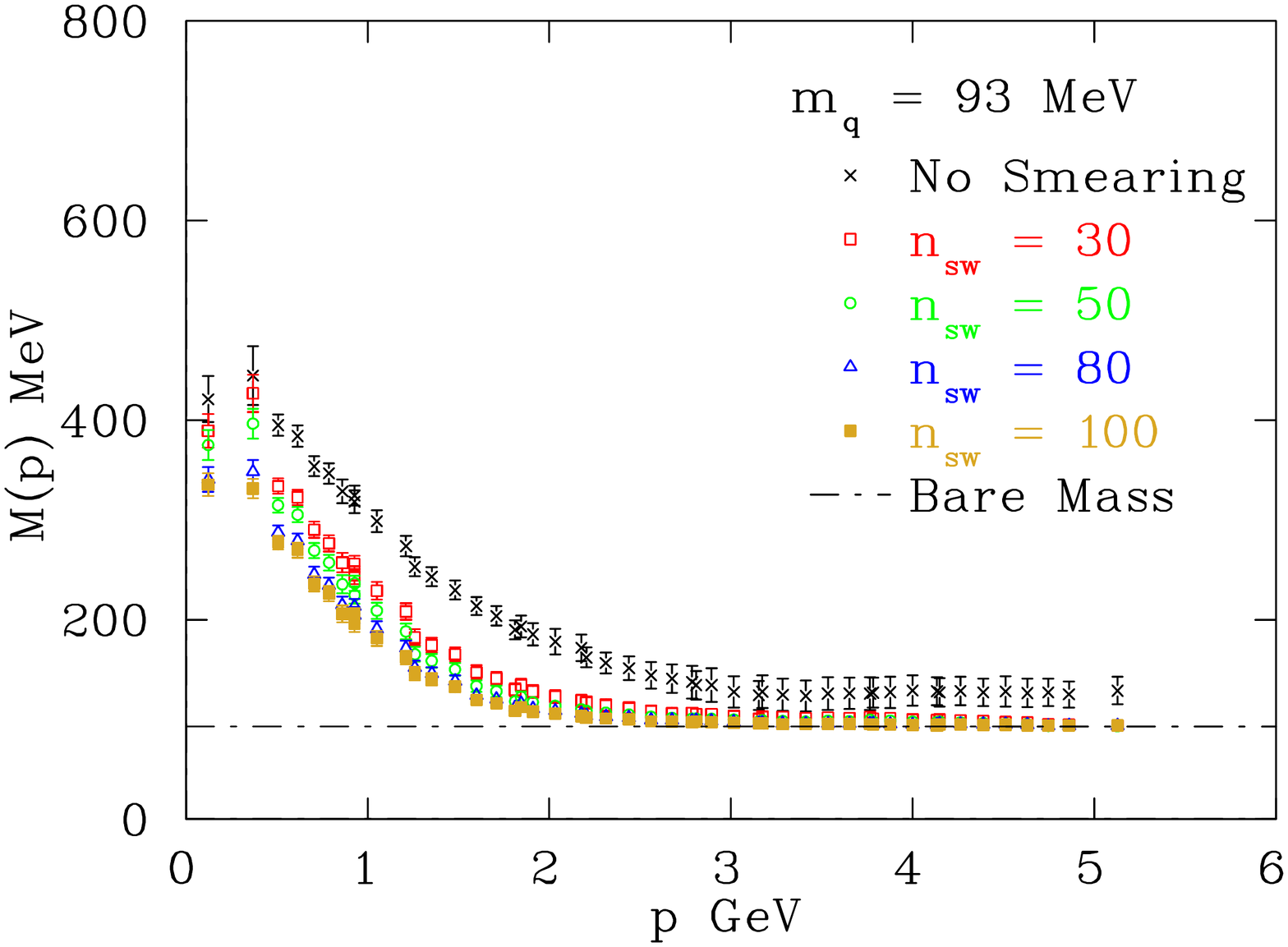}
\includegraphics[trim=1cm 2cm 2cm 2.5cm, clip=true,scale=0.26,angle=90]{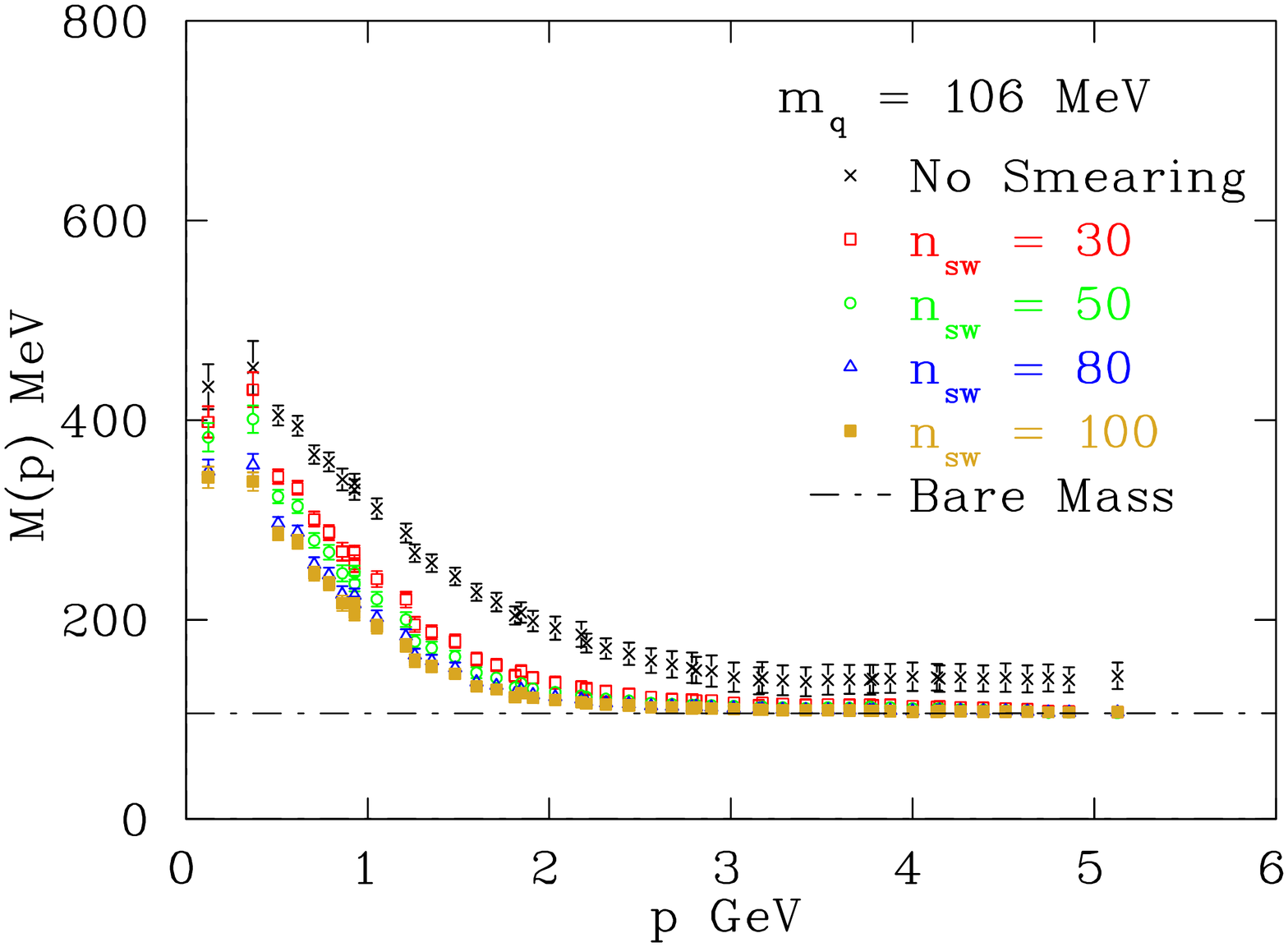}
\includegraphics[trim=1cm 2cm 2cm 2.5cm, clip=true,scale=0.26,angle=90]{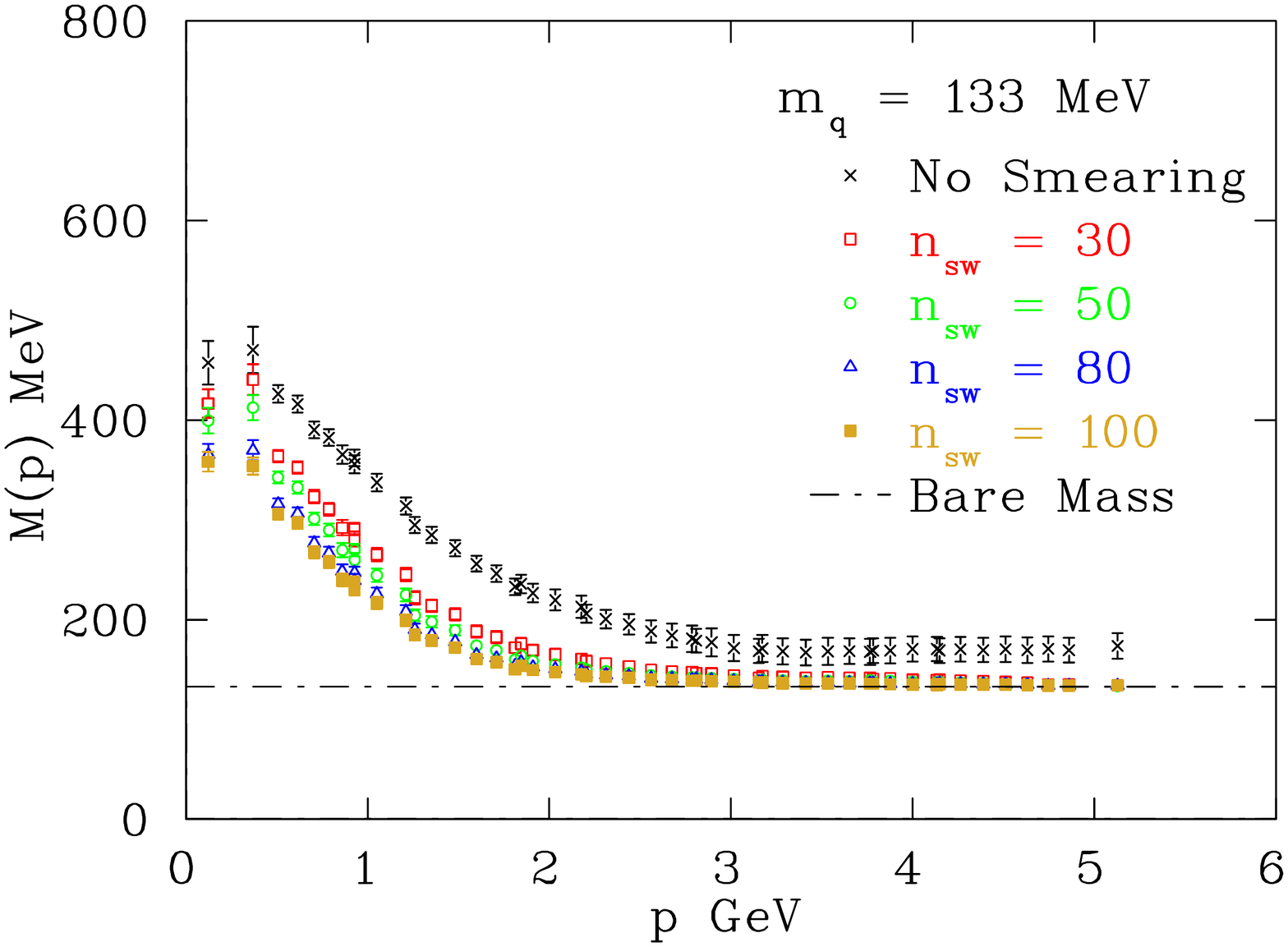}
\includegraphics[trim=1cm 2cm 2cm 2.5cm, clip=true,scale=0.26,angle=90]{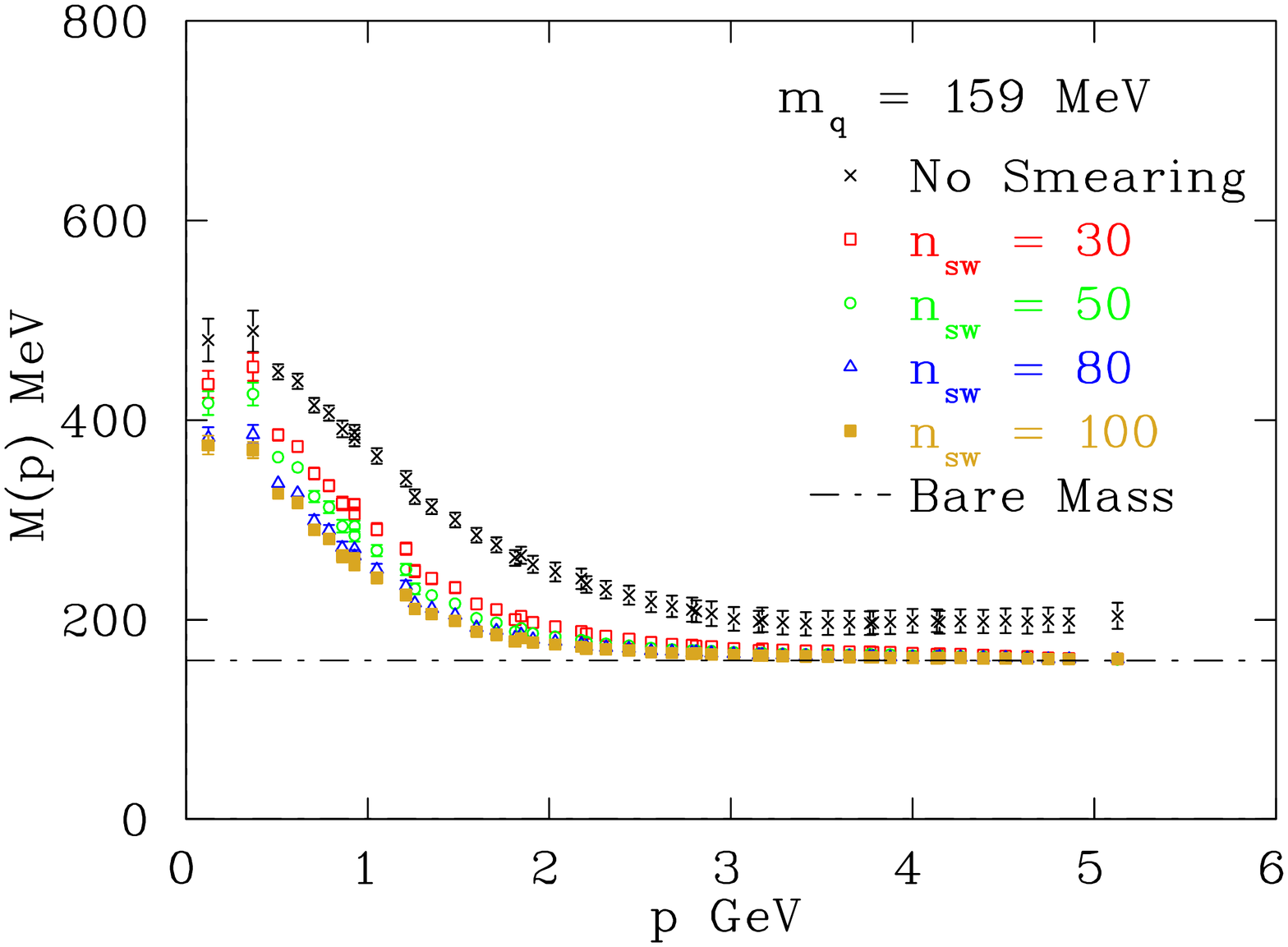}
\includegraphics[trim=1cm 2cm 2cm 2.5cm, clip=true,scale=0.26,angle=90]{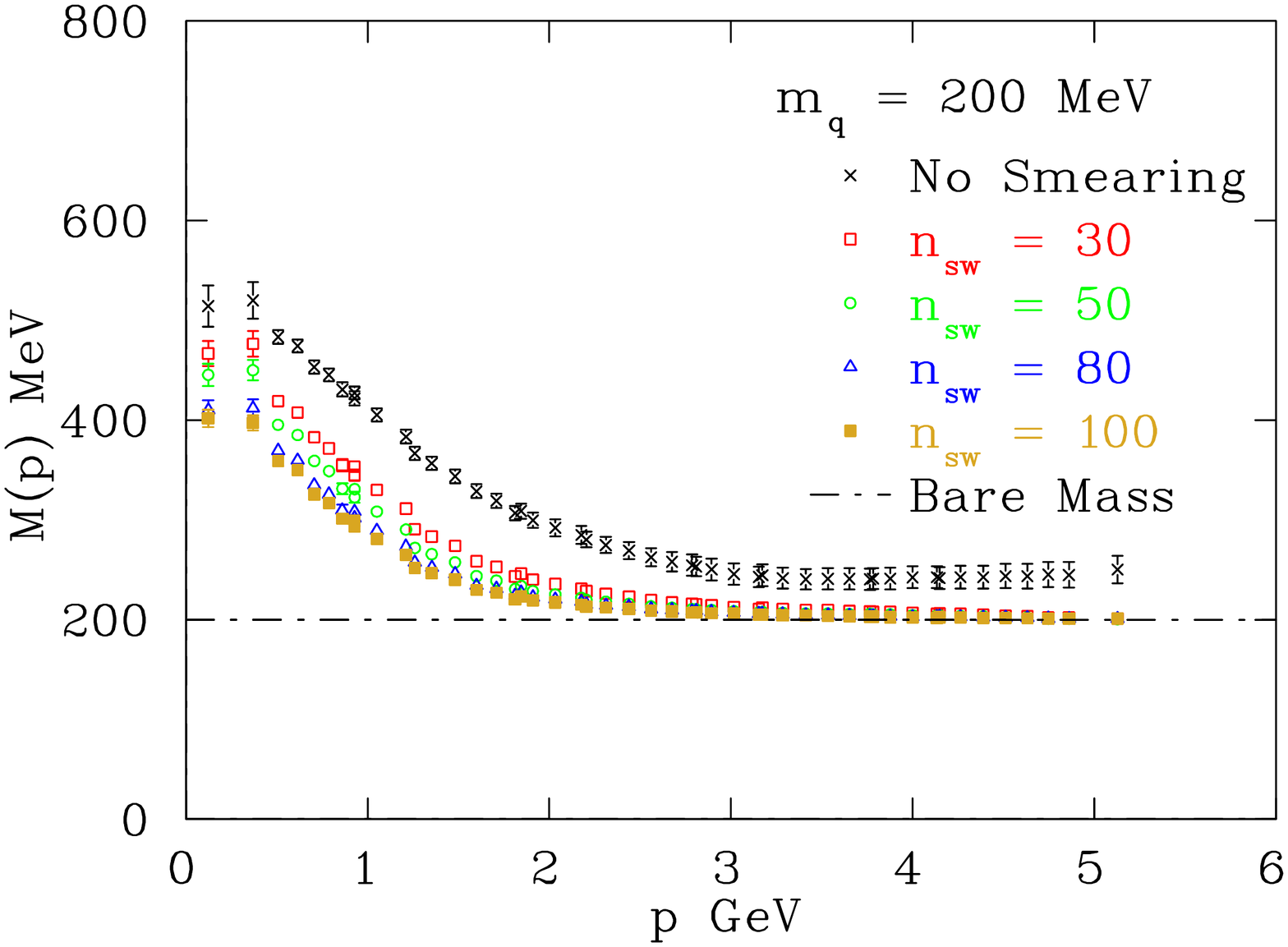}
\caption{The mass function at various values of bare quark mass, indicated by the dot-dash line. The same value of bare mass is used for both smeared and unsmeared results. Configurations with 0,30, 50, 80 and 100 sweeps of smearing are considered.}
\label{fig:Mhigh}
\end{figure*}

Higher masses are shown in Fig.~\ref{fig:Mhigh} for all $p$, using the same bare mass for both smeared and unsmeared configurations. These show broadly similar results at all masses, though the gap between levels of smearing becomes clearer for larger masses, which should be more sensitive to the disruption of short distance physics. This reinforces the necessity of maintaining a high density of topological objects in the QCD vacuum and reinforcing the danger of destroying topological objects with excess smearing. The largest gap, however, remains between no smearing and 30 sweeps.

\section{Conclusion}
\label{sec:conclusion}

We have used over-improved stout-link smearing, to reveal an underlying gauge field structure resembling an instanton liquid. After around 50 sweeps we can be confident the lattice is dominated by instanton-like objects.\par 
Our calculations of the non-perturbative mass function on smeared configurations reveals that it retains its qualitative shape at even high levels of smearing. There is some loss of dynamical mass generation, increasing with smearing, which can be attributed to a thinning of the vacuum through the destruction of instanton/anti-instanton pairs by the smearing algorithm. Regardless, we have shown that a gauge configuration consisting solely of instanton-like objects can accurately reproduce the majority of the long-range behaviour of the quark propagator, and thus conclude that instantons are the primary mechanism responsible for the dynamical generation of mass.

\section*{Acknowledgments}
\label{sec:acknowledgments}

We wish to thank Craig Roberts for discussions inspiring this quantitative evaluation of the role of QCD instantons in the quark propagator. We also thank Peter Moran for his input. This research was undertaken with the assistance of resources at the NCI National Facility in Canberra, Australia, and the iVEC facilities at Murdoch University (iVEC@Murdoch) and the University of Western Australia (iVEC@UWA). These resources were provided through the National Computational Merit Allocation Scheme, supported by the Australian Government. This research is supported by the Australian Research Council.

\end{document}